\documentclass[11pt, a4paper]{article}
\pdfoutput=1

\usepackage{graphicx}
\usepackage{xcolor}
\usepackage{amsmath, amsthm, tikz}
\usepackage{float}

\usepackage[style=authoryear, giveninits=true, isbn=false, uniquelist=false, terseinits=true, uniquename=false, mincitenames = 1, maxcitenames=2, maxbibnames=99]{biblatex}
\bibliography{reference}
\DeclareNameAlias{sortname}{family-given}
\AtEveryBibitem{\clearfield{labelmonth}}
\AtEveryCitekey{\clearfield{labelmonth}}
\AtEveryBibitem{\clearfield{extradate}}
\renewbibmacro*{volume+number+eid}{%
  \printfield{volume}%
  \printfield{number}%
  \setunit{\addcomma\space}%
  \printfield{eid}}
\DeclareFieldFormat[article]{number}{\mkbibparens{#1}}

\DefineBibliographyStrings{english}{%
  page             = {\ifbibliography{}{}},
  pages            = {\ifbibliography{}{}},
}
\renewbibmacro{in:}{}

\usepackage{authblk}
\usepackage{comment}
\usepackage{soul}
\usepackage{amssymb}
\usepackage{algorithmicx}
\usepackage{algorithm}
\usepackage{algpseudocode}
\usepackage{microtype}

\RequirePackage{amsfonts}
\usepackage{mathtools}
\usepackage{arydshln}  

\usepackage{comment}
\newcommand{\thmcrossref}[1]{#1}

\usepackage[cm, empty]{fullpage}

\definecolor{mygreen}{RGB}{73, 157, 94}
\usepackage{mathrsfs} 

\graphicspath{{Fig/}}

\theoremstyle{thmstyleone}%
\newtheorem{theorem}{Theorem}

\newtheorem{lemma}{Lemma}
\newtheorem{RULE}{Rule}

\theoremstyle{thmstyletwo}%
\newtheorem{example}{Example}%
\theoremstyle{thmstylethree}%
\newtheorem{definition}{Definition}

\newcommand{\indep}{\,\raisebox{0.05em}{\rotatebox[origin=c]{90}{$\models$}}\,}

\newcommand{\nindep}{\,\raisebox{0.05em}{$\not\!\perp\!\!\!\perp$}}

\newcommand{\given}{\mid}
\usetikzlibrary{decorations.text,fit}

\newcommand{\appropto}{\mathrel{\vcenter{
  \offinterlineskip\halign{\hfil$##$\cr
    \propto\cr\noalign{\kern2pt}\sim\cr\noalign{\kern-2pt}}}}}

\makeatletter
\newcommand\approxsim{\mathchoice
  {\@approxsim {\displaystyle}      {1ex} }
  {\@approxsim {\textstyle}         {1ex} }
  {\@approxsim {\scriptstyle}       {.7ex}}
  {\@approxsim {\scriptscriptstyle} {.5ex}}}
\newcommand\@approxsim[2]{%
  \mathrel{%
    \ooalign{%
      $\m@th#1\sim$\cr
      \hidewidth$\m@th#1.$\hidewidth\cr
      \hidewidth\raise #2 \hbox{$\m@th#1.$}\hidewidth\cr
    }%
  }%
}
\makeatother

\DeclareMathOperator*{\argmin}{arg\,min}

\makeatletter
\@addtoreset{section}{part}
\title{A General Framework for Cutting Feedback within Modularised Bayesian Inference}

\author[1]{Yang Liu\textsuperscript{$\ast$}}
\author[1]{Robert J.B. Goudie}
\affil[1]{MRC Biostatistics Unit, University of Cambridge, UK}

\begin{document}





\maketitle
\abstract{Standard Bayesian inference enables building models that combine information from various sources, but this inference may not be reliable if components of the model are misspecified. Cut inference, a particular type of modularised Bayesian inference, is an alternative that splits a model into modules and cuts the feedback from any suspect module. Previous studies have focused on a two-module case, but a more general definition of a ``module'' remains unclear. We present a formal definition of a ``module'' and discuss its properties. We formulate methods for identifying modules; determining the order of modules; and building the cut distribution that should be used for cut inference within an arbitrary directed acyclic graph structure. We justify the cut distribution by showing that it not only cuts the feedback but also is the best approximation to the joint distribution satisfying this condition in Kullback-Leibler divergence. We also extend cut inference for the two-module case to a general multiple-module case via a sequential splitting technique and demonstrate this via illustrative applications.}

\section{Introduction}
Standard Bayesian inference is a powerful tool when one believes the whole model to be correctly specified, but requiring every piece of a model to be correctly specified is often unrealistic. Several robust Bayesian inference methods have been established when the whole model is misspecified. One class of approaches adopts a tempered likelihood where the likelihood is raised to a power between 0 and 1, leading to a power posterior or fractional posterior, and a Bayesian update is conducted thereafter \parencite[e.g.,][]{https://doi.org/10.1111/j.1467-9868.2007.00650.x,https://doi.org/10.1111/rssb.12158,10.1093/biomet/asx010, bhattacharya2019, doi:10.1080/01621459.2018.1469995}. Another class of approaches replaces the distribution of the likelihood with heavy-tailed distributions, for example via individual-specific variance parameters, to account for conflicting information sources \parencite[e.g.,][]{10.1214/11-BJPS164, doi:10.1080/03610926.2011.592250, 10.1214/17-BA1090}. However, especially when dealing with a complex true generating process, it is impossible to allocate equal confidence to all aspects of the model. An alternative approach may be desirable in which the contribution from each aspect of the model to inference of shared parameters can vary so that unreliable aspects are confined and minor misspecifications do not affect the whole model. This leads to the development of modularised Bayesian inference where one model is divided into ``modules'' such that the handling of partial misspecification is possible \parencite{lunn2009combining,liu2009modularization}.

One type of modularised Bayesian inference is cut inference, which  ``cuts the feedback'' to manipulate ``influence'' among modules. By influence, we mean the flow of information that affects the estimation of particular parameters. Within a Bayesian framework, parameters are usually influenced by information from more than one module. The purpose of cut inference is to prevent suspect modules from influencing reliable modules. In a simple two-module case where misspecification exists in only one suspect module, we can estimate the reliable module solely within a standard Bayesian framework and then estimate the suspect module by conditioning on everything that is either known or inferred by the reliable module. Note that this leads to a cut distribution rather than a standard posterior distribution because estimation of the reliable module is not influenced by the suspect module, as it would be under standard Bayesian inference. The cut distribution is normally intractable and is difficult to sample from using Monte Carlo sampling methods. The samplers implemented in WinBUGS \parencite{https://doi.org/10.1002/sim.3680} may not have the cut distribution as their stationary distribution \parencite{plummer2015cuts}. Several alternative sampling methods have been proposed \parencite{https://doi.org/10.1111/rssb.12336, yu2021variational, pompe2021asymptotics, liu2020stochastic}. An alternative to cut inference is to reduce the influence of suspect modules, rather than completely prevent it. This leads to the proposal of semi-modular inference model \parencite{carmona2020semi,liu2021generalized,nicholls2022valid}.

In this paper, we consider the fundamental nature of modularised Bayesian inference and propose a general framework for cut inference in general Bayesian statistical models whose joint distribution satisfies the Markov factorization property with respect to a directed acyclic graph (DAG). Although modularised Bayesian inference has been applied in various areas, studies of its methodology, theory and algorithms are so far based on a specific two-module case in which observations $y$ depend on parameters $\varphi$, and $\theta$ and observations $z$ are solely dependent upon $\varphi$. Going beyond this simple setting requires considering several problems. First, a fundamental question is how to define a module because the definition of a ``module'' remains unclear in the literature. Second, how can one formally identify influence among modules and implement cut inference under this more complicated structure? Specifically, we aim to answer the following three questions in concise mathematical language: given an arbitrary design of a model, (1) how to define modules and how to analyze them via a graphical model; (2) how to identify influence among modules; and (3) how to cut the feedback. All proofs are collated in Appendix A in the supplementary materials.

\section{Methods}
We consider observable random variables $x$, which can include fixed, known parameters as well as observable quantities; and parameters $\theta$, which can represent unobserved data as well as standard parameters. We denote the set of $n$ observable random variables $X=(x_1,x_2,\dots,x_n)$, which are not necessarily identically or independently distributed, and the set of $m$ parameters $\Theta=(\theta_1,\theta_2,\dots,\theta_m)$.
We aim to obtain the true data generating process $\check{p}(X)$ but usually this is unattainable.
Instead, we specify a model that we hope is flexible enough to describe the true data generating process of $X$. However, as we mentioned above, realistically there is often either partial misspecification; or some of the observations are regarded as more reliable than others. A natural solution is to partition the observations into several groups and analyze them to some degree separately, as proposed by modularised Bayesian inference.

We will assume that one subset of observable random variables $X^\ast\subseteq X$ is of primary interest within one ``module'' (which we will formally define later as a ``self-contained Bayesian module'') and that we seek to estimate its true data generating process. To do this, we build a $X^\ast$-associated model, which can involve both other observable random variables (known) and parameters (unknown). To infer these associated unknown parameters within a Bayesian framework, we also consider a prior model, which can depend on additional observable random variables (known).

After introducing notation in the next section, we formalize these notions to define a ``module''. We will also need to consider the internal relationships and relative priorities of several modules, which together describe all observable random variables $X$. This leads to consideration of the ``ordering'' of modules. We consider the identification of module ordering and formulate the cut distribution within the 
two- and three-module cases. Based on these basic cases, we extend the framework to cover more than three modules via a sequential splitting technique.

\subsection{Notation and definitions}
We assume the set of all variables $\Psi=(X,\Theta)$ has a joint distribution such that densities exist for all variables. We further assume the joint distribution satisfies the Markov factorization property with respect to a DAG $G=(\Psi,\mathcal{E})$, where each node is associated with a component of $\Psi$ and each directed edge $(\psi_1, \psi_2)\in \mathcal{E}\subseteq\Psi\times\Psi$. We will use $\psi_1\rightarrow\psi_2$ and $\psi_2\leftarrow\psi_1$ to denote a directed edge $(\psi_1, \psi_2)\in \mathcal{E}$. The DAG implies the joint distribution of $\Psi$ can be factorized as
\begin{equation}
p(\Psi)=p(X,\Theta)=\prod_{1\leq i\leq n,\,1\leq j\leq m}p\left(x_i\given\text{pa}(x_i))\,p(\theta_j\given \text{pa}(\theta_j)\right),
\label{E8}
\end{equation}
where $\text{pa}(\psi)=\{a\in\Psi\ \given a\rightarrow \psi \} $ is the set of parent nodes of node $\psi \in \Psi$. When $\text{pa}(\psi)=\emptyset$, then we assume that $p(\psi)$ is a fixed, known distribution that does not depend on any component of $\Psi$. We denote the set of parent nodes $\text{pa}(B)=\{a\in\Psi\setminus B\ \given a\rightarrow b_i \in B\}$ of the set $B=(b_1,\dots,b_q)$. Note that under our definition $b_i\notin \text{pa}(B)$ for all $i=1,\dots,q$. Denote $\text{ch}(\psi)=\{a\in\Psi\ \given a\leftarrow \psi \} $ as the set of child nodes of node $\psi \in \Psi$. Similarly, we denote the set of child nodes $\text{ch}(B)=\{a\in\Psi\setminus B\ \given a\leftarrow b_i \in B\}$ of the set $B=(b_1,\dots,b_q)$. Note that $b_i\notin \text{ch}(B)$ for all $i=1,\dots,q$.

We say that there is a directed path $a \rightsquigarrow b$ with path $a=\psi_1,\psi_2,\dots,\psi_s=b$ between two distinct nodes $a$ and $b$ in the DAG if $(\psi_i, \psi_{i+1}) \in \mathcal{E}$ for all $1\leq i\leq s - 1$. We say $a$ is the root of this path and $b$ is the leaf of this path. In addition, we say that $a$ is an ancestor of $b$, and we denote $\text{an}(b) = \{a\in\Psi\ \given a \rightsquigarrow b\}$ as the set of ancestors of $b$. Note that $b\notin \text{an}(b)$. Similarly, we denote the set of ancestors $\text{an}(B) = \{a\in\Psi\setminus B\ \given a \rightsquigarrow b_i \text{ for some } b_i \in B\}$ of the set $B=(b_1,\dots,b_q)$. Note that $b_i\notin \text{an}(B)$ for all $i=1,\dots,q$.

\subsection{Self-contained Bayesian modules}\label{SE 2.2}
Our definition of a self-contained Bayesian module is described in terms of the variables that are involved in the $X^\ast$-associated model and prior model when $X^\ast\subseteq X$ is the primary focus in a given module, as given by
the following definition:
\begin{definition}[$X^\ast$-associated parameters and observables]
\ Consider variables $\Psi=(X,\Theta)$ with joint distribution satisfying the Markov factorization property with respect to a DAG $G=(\Psi,\mathcal{E})$, and observable random variables $X^\ast\subseteq X$, whose true generating process is of interest. Then denote
\begin{itemize}
\item The observable ancestors of $X^\ast$ by $\text{an}^X(X^\ast) := \text{an}(X^\ast)\cap X$.
\item The $X^\ast$-associated parameters as the parameter ancestors of $X^\ast$ that are not conditionally-independent of $X^\ast$ given its observable ancestors:
\[
\Theta_{X^\ast} := \left\{\theta\in \text{an}(X^\ast)\cap\Theta\ \middle|\ \left(\theta \nindep X^\ast \given  \text{an}^X(X^\ast) \right)\right\}.
\]
\item The $X^\ast$-associated observables as the observable ancestors of $X^\ast$ that are not conditionally-independent of $X^\ast$ given its other observable ancestors:
\[
X_{X^\ast} := \left\{x\ \in \text{an}^X(X^\ast) \ \middle|\ \left((x \nindep X^\ast \given  \text{an}^X(X^\ast)\setminus x)\right)\right\}.
\]
\item All $X^\ast$-associated variables by $\Psi_{X^\ast}=(X_{X^\ast}, \Theta_{X^\ast})$.
\end{itemize}
\end{definition}

\begin{example}
To illustrate our methods, we consider a generic epidemiological model with measurement error \parencite{richardsonConditionalIndependenceModels1993}, represented by the DAG in Figure \ref{F1}. ``Two-stage'' methods for cut distributions have been previously proposed for similar epidemiological problems \parencite[e.g.][]{blangiardoTwostageBayesianModel2016}.
The aim is to estimate the association, quantified by the parameter $\beta$, between the exposure $W$ and the outcome $Y$ and confounders $C$, with interest primarily on the relationship with $W$.
However, the main study data record for individuals $i = 1, \dots, m$ only a surrogate $Z_i$ for exposure, rather than exposure $W_i$ itself.
These data are supplemented by observations $i=m+1, \dots, n$ from an external validation study that records the surrogate $Z_i$, exposure $W_i$ and confounders $C_i$ but not the outcome $Y_i$.
We suppose, as would be common, that the external validation study has small sample size, with $n-m \ll m$, but the data have been collected more carefully and systematically than the main study data.
For this reason, we assume the validation study provides more reliable information about the relationships between $C$, $W$ and $Z$ than the main study data, which motivates a modularised approach.

\begin{figure}[t]
\setlength{\abovecaptionskip}{0cm}
\setlength{\belowcaptionskip}{0cm}
\centering
\begin{tikzpicture}[minimum width=0.75cm, minimum height = 0.75cm]
\node[draw] (Wi) at (0.5, 3) {$W_i$};
\node[draw] (Ci) at (2, 4.5) {$C_i$};
\node[draw] (Zi) at (-1, 0) {$Z_i$};

\node[circle, draw] (beta) at (-8, 1.5) {$\beta$};
\node[circle, draw] (lambda) at (-2.5, 1.5) {$\lambda$};
\node[circle, draw] (pi) at (-2.5, 4) {$\pi$};

\node[anchor=west] (yplate) at (-1.75, -0.75) {$i = m+1, \dots, n$};
\draw[rounded corners] (2.5, -1) rectangle (-1.75, 5);

\node[draw] (Yj) at (-5.5, 0) {$Y_i$};
\node[circle, draw] (Wj) at (-5.5, 3) {$W_i$};
\node[draw] (Cj) at (-7, 4.5) {$C_i$};
\node[draw] (Zj) at (-4, 0) {$Z_i$};

\node[anchor=north west] (label1) at (-1.75, 5.15) {External validation};
\node[anchor=north east] (label1) at (-3.25, 5.15) {Main study};

\node[anchor=west] (yplate) at (-7.5, -0.75) {$i = 1, \dots, m$};
\draw[rounded corners] (-3.25, -1) rectangle (-7.5, 5);

\draw[dashed] (-8.5,-1.25) -- (-4.75,-1.25) -- (-4.75,2.5) -- (-2,2.5) -- (-2,5.25) -- (-8.5,5.25) -- cycle;

\draw[->] (Ci) -- (Wi);
\draw[->] (Wi) -- (Zi);

\draw[->] (Wj) -- (Yj);
\draw[->] (Cj) -- (Yj);
\draw[->] (Cj) -- (Wj);
\draw[->] (Wj) -- (Zj);

\draw[->] (pi) -- (Wi);
\draw[->] (lambda) -- (Zi);

\draw[->] (beta) -- (Yj);
\draw[->] (pi) -- (Wj);
\draw[->] (lambda) -- (Zj);

\end{tikzpicture}
\caption[Self-contained Bayesian module.]{\textbf{Self-contained Bayesian module.} Squares denote observable random variables and circles denote parameters. The dashed part is a minimally self-contained Bayesian module for $Y_{1:m}$ (see Definition \ref{DE1}).}
\label{F1}
\end{figure}
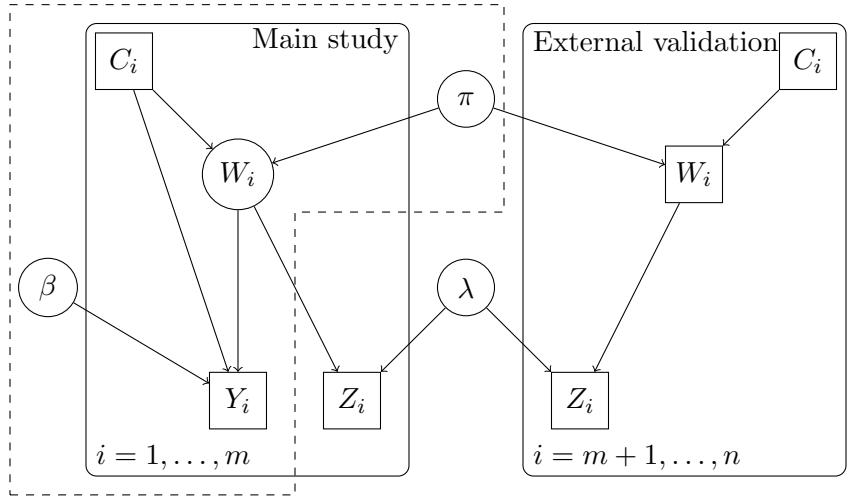

Initially, consider only the part of the model enclosed by a dashed line.
The data are $X = \{Y_{1:m}, C_{1:m}\}$ and the parameters are $\Theta = \{\beta, W_{1:m}, \pi\}$.
The primary focus is the outcome $X^\ast = Y_{1:m}$.
The observable ancestors $\text{an}^X(X^\ast) = C_{1:m}$, the $X^\ast$-associated parameters $\Theta_{X^\ast} = \{\beta, W_{1:m}, \pi\}$, and $X_{X^\ast} = \{C_{1:m}\}$.
Here $\prod_{i=1}^{m}p(Y_i \given \beta, W_i, C_i)$ is the $X^\ast$-associated model and $p(\beta)\,p(\pi)\prod_{i=1}^{m}p(W_i \given \pi, C_i)$ is the prior model. Observe that the posterior distribution for $\Theta_{X^\ast}$ given $X^\ast$ and $X_{X^\ast}$ can be inferred within a standard Bayesian framework as
\begin{equation}
\begin{split}
\label{E9}
p(\Theta_{X^\ast} \given X^\ast, X_{X^\ast}) &=
p(\beta \given Y_{1:m}, C_{1:m})\\
&\propto p(\beta)
\int
p(\pi)
\prod_{i=1}^{m}
p(Y_i \given \beta, W_i, C_i)\,p(W_i \given \pi, C_i)
dW_{1:m} d\pi
\end{split}
\end{equation}
Note that, since our focus is on $Y_{1:m}$, we would not need to consider any parameters $\theta \in \text{an}(C_{1:m})$ that are conditionally independent of $X^\ast = Y_{1:m}$ in \eqref{E9}.
\end{example}

Often there is a degree of choice about which data are used when inferring the parameter of interest $\Theta_{X^\ast}$. Specifically, there may be support variables $\Psi_{\text{supp}}$ that can be added to the existing variables $X^\ast$ and $X_{X^\ast}$ to form a different model involving $\Theta_{X^\ast}$.
Clearly, the posterior distribution for $\Theta_{X^\ast}$ may be altered by including $\Psi_{\text{supp}}$.
We formalise this idea through the following definition.
\begin{definition}[Self-contained Bayesian module for $X^\ast$]
\label{DE1}
Consider variables $\Psi=(X,\Theta)$ with joint distribution satisfying the Markov factorization property with respect to a DAG $G=(\Psi,\mathcal{E})$; observable random variables $X^\ast$, whose true generating process is of interest, with $X^\ast$-associated parameters $\Theta_{X^\ast}$ and $X^\ast$-associated observables $X_{X^\ast}$; and support variables $\Psi_{\text{supp}}=(X_{\text{supp}},\Theta_{\text{supp}}) \subseteq \Psi \setminus (\Psi_{X^\ast}\cup X^\ast)$. We say that a set of variables $(X^\ast,\Psi_{X^\ast},\Psi_{\text{supp}})$ form a self-contained Bayesian module for $X^\ast$ that can be used to estimate the true data generating process of $X^\ast$ if the posterior distribution given by
\[
p(\Theta_{X^\ast}\given X^\ast,X_{X^\ast},X_{\text{supp}})=\int p(\Theta_{X^\ast},\Theta_{\text{supp}}\given X^\ast,X_{X^\ast},X_{\text{supp}})d\Theta_{\text{supp}}
\]
is well defined given the DAG. When $\Psi_{\text{supp}}=\emptyset$, we say $(X^\ast,\Psi_{X^\ast})$ forms a minimally self-contained Bayesian module for $X^\ast$.
\end{definition}

\addtocounter{example}{-1}
\begin{example}[continued]
To better understand Definition \ref{DE1}, consider the whole illustrative example in Figure \ref{F1}, not just that enclosed by a dashed line.
The additional components are the surrogate exposure $Z_{1:m}$ for the main study participants, and $n-m$ observations from the external validation study.
The \textit{support observed variables} $X_{\text{supp}} = \{Z_{1:m}, Z_{(m+1):n}, W_{(m+1):n}, C_{(m+1):n}\}$
and \textit{support parameters} $\Theta_{\text{supp}} = \{\lambda\}$ indirectly provide information about the parameter of interest $\beta$, via the \textit{support models} $\prod_{i=1}^{m}
p(Z_i \given \lambda, W_i)$ and $\prod_{i=m+1}^{n}
p(Z_i \given \lambda, W_i)\,
p(W_i \given \pi, C_i)$.

Combining the support model with the $X^{\ast}$-associated model, an alternative posterior distribution is given by
\begin{multline}
\label{E10}
p(\Theta_{X^\ast}\given X^\ast,X_{X^\ast},X_{\text{supp}}) =
p(\beta \given Y_{1:m}, W_{(m+1):n}, C_{1:n}, Z_{1:n})\\
\propto
p(\beta)
\int
p(\pi)\,p(\lambda)
\prod_{i=1}^{m}
p(Y_i \given \beta, W_i, C_i)\,
p(Z_i \given \lambda, W_i)\,
p(W_i \given \pi, C_i)
\\
\times \prod_{i=m+1}^{n}
p(Z_i \given \lambda, W_i)\,
p(W_i \given \pi, C_i)
dW_{1:m} d\pi d\lambda,
\end{multline}
Both \eqref{E9} and \eqref{E10} can be used to infer the parameter of interest $\Theta_{X^\ast} = \beta$ via a standard Bayesian framework and subsequently estimate the true data generating process of $X^\ast$. The difference is that \eqref{E9} involves only the minimum information that is required to infer $\Theta_{X^\ast}$ whereas \eqref{E10} involves additional information.
\end{example}

Self-contained Bayesian modules have two appealing properties. First, the module is defined with respect to a set of observable random variables $X^\ast$. Hence, the inference of the parameter $\Theta_{X^\ast}$ is meaningful in the sense that it can be always used to infer the data generating process of $X^\ast$. Second, clearly we can conduct standard Bayesian inference using only information from the self-contained module. This makes it possible to prevent information from outside of the module affecting inference because the module contains sufficient information to infer $\Theta_{X^\ast}$ by itself within the Bayesian framework. This property will be helpful for modularised Bayesian inference.

\subsection{Two-module case}
\label{Sec2}
For the two-module case, we first consider how to form two self-contained Bayesian modules. We then derive the standard Bayesian posterior in terms of these two modules, before considering cut distributions. Specifying cut distributions will require us to identify the ordering of the two modules. We then justify the cut distribution in this setting.

\subsubsection{Constructing self-contained modules}
To group variables into two
self-contained Bayesian modules, we start by partitioning the observable random variables $X$ into two disjoint sets. We call these sets $X_A^\ast$ and $X_B^\ast$. We then, according to Rule~\ref{Ar1} (below), enlarge these sets with additional variables until both sets form (potentially overlapping) modules that can be inferred without using information that exists exclusively in the other module. In other words, we enlarge the sets until both are minimal self-contained Bayesian modules, as we will prove in Theorem~\ref{l3}. It is important to note that the two modules are often not disjoint: modules overlap whenever there are quantities without which both modules are incomplete.
\begin{RULE}[Constructing self-contained modules]
\label{Ar1}
Given a DAG $G=(\Psi,\mathcal{E})$ with $\Psi=(X,\Theta)$ and a set of observables $X_A^\ast \subseteq X$, we form the corresponding module set $\Psi_A = (X_A,\Theta_A)$ by first setting $X_A = X_A^\ast$ and then adding further variables to $X_A$ and $\Theta_A$ as follows. For every directed path $b \rightsquigarrow a$ with leaf $a\in X_A^\ast$ in the DAG $G$, with path components labelled $a=\psi_1,\psi_2,\dots,\psi_s=b$, we consider the following two cases:
\begin{enumerate}
    \item If there exists an observable $\psi_r\in X \setminus X_A^\ast$, $r>1$ such that $(\psi_1,\dots,\psi_{r-1})\notin X \setminus X_A^\ast$, then we add into $\Psi_A$ the initial path components $(\psi_1,\psi_2,\dots,\psi_r)$.
    \item If the directed path $b \rightsquigarrow a$ does not involve any node from $X \setminus X_A^\ast$, then we add into $\Psi_A$ the complete path $(\psi_1,\psi_2,\dots,\psi_s)$.
\end{enumerate}
\end{RULE}

Using Rule~\ref{Ar1}, we form two modules $\Psi_A=(X_A,\Theta_A)$, which we call module A, and $\Psi_B=(X_B,\Theta_B)$, which we call module B. 
We interchangeably use module $I$ to refer to the set $\Psi_I$ and use $\Psi_I$ to refer to the module $I$, where $I$ is an arbitrary index.

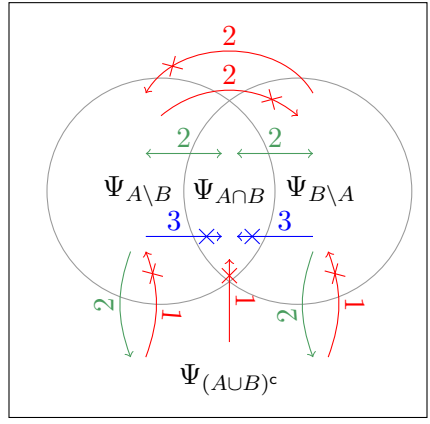
\begin{figure}[t]
\centering
\tikzset{pics/.cd,
    Cross/.style args={#1 and #2}{%
        code = {%
        \draw[#2,rotate=45,scale=1]
                (0,#1 pt) -- (0,-#1 pt) ;
        \draw[#2,rotate=-45,scale=1]
                (0,#1 pt) -- (0,-#1 pt) ;
        }
    },
    Cross/.default={2.5 and gray!25!black}}
\begin{tikzpicture}[x=1cm, y=1cm]
\node[draw, black!40, circle, minimum size =3cm, anchor = center] (circle1) at (0,0){};
\node[draw, black!40, circle, minimum size =3cm, anchor = center] (circle2) at (1.8,0){};

\draw (-2,-3) rectangle (3.5,2.5);

\node at (-0.3,0) {$\Psi_{A\setminus B}$};
\node at (0.9,0) {$\Psi_{A\cap B}$};
\node at (2.1,0) {$\Psi_{B\setminus A}$};
\node at (0.9,-2.5) {$\Psi_{(A\cup B)^\mathsf{c}}$};

\draw[red, ->,postaction={decorate,decoration={text color=red, text={\space\space 1},reverse path,text along path,text align=center,raise=0.5ex}}] (-0.2, -2.2) to [bend right=20]  pic[pos=.8,-,rotate=30] {Cross={4 and red}} (-0.2,-0.8);
\draw[mygreen, ->,postaction={decorate,decoration={text color=mygreen, text={2},reverse path, text along path,text align=center,raise=0.5ex}}] (-0.4, -0.8) to [bend right=20] (-0.4,-2.2);

\draw[red, ->,postaction={decorate,decoration={text color=red, text={1},reverse path,text along path,text align=center,raise=0.5ex}}] (2.2, -2.2) to [bend right=20]  pic[pos=.8,-,rotate=30] {Cross={4 and red}} (2.2,-0.8);
\draw[mygreen, ->,postaction={decorate,decoration={text color=mygreen, text={2\space\space},reverse path, text along path,text align=center,raise=0.5ex}}] (2, -0.8) to [bend right=20] (2,-2.2);

\draw[red, ->,postaction={decorate,decoration={text color=red, text={2}, text along path,text align=center,raise=0.5ex}}]  (0, 1) to [bend left=40] pic[pos=.8,-,rotate=-30] {Cross={4 and red}} (1.8,1);
\draw[red, ->,postaction={decorate,decoration={text color=red, text={2},reverse path, text along path,text align=center,raise=0.5ex}}] (2, 1.3) to [bend right=60] pic[pos=.8,-,rotate=30] {Cross={4 and red}} (-0.2,1.3);

\draw[red, ->,postaction={decorate,decoration={text color=red, text={1},reverse path, text along path,text align=center,raise=0.5ex}}] (0.9, -2) to pic[pos=.8,-] {Cross={4 and red}} (0.9,-0.9);

\draw[mygreen, <->,postaction={decorate,decoration={text color=mygreen, text={2},reverse path, text along path,text align=center,raise=0.5ex}}] (0.8, 0.5) to (-0.2,0.5);
\draw[mygreen, <->,postaction={decorate,decoration={text color=mygreen, text={2}, text along path,text align=center,raise=0.5ex}}] (1, 0.5) to (2,0.5);

\draw[blue, ->,postaction={decorate,decoration={text color=blue, text={3\space\space},text along path,text align=center,raise=0.5ex}}] (-0.2, -0.6) to pic[pos=.8,-] {Cross={4 and blue}} (0.8,-0.6);
\draw[blue, ->,postaction={decorate,decoration={text color=blue, text={\space\space 3},reverse path, text along path,text align=center,raise=0.5ex}}] (2, -0.6) to pic[pos=.8,-] {Cross={4 and blue}} (1,-0.6) ;

\end{tikzpicture}
\caption[Venn diagram illustrating the partition of $\Psi$ and the links between these sets]{\textbf{Venn diagram illustrating the partition of $\Psi$ and the links between these sets.} Arrows (in green) indicate links that may exist, whereas crossed arrows (in red) indicate links that cannot exist. The crossed links (in blue) indicate a v-structure that cannot exist. The numbers on links indicate the part of Lemma~\ref{l1} that proves the corresponding property.}
\label{F2}
\end{figure}

To formulate posterior and cut distributions involving these modules, we will need the following partition of $\Psi$. Let $\Psi_{A\setminus B}$ denote the set of variables that belong to module $A$ but not module $B$, and let $\Psi_{B\setminus A}$ be defined correspondingly. Furthermore let $\Psi_{A\cap B}$ denote the set of variables that belong to both modules, $\Psi_{A\cup B}$ denote the set of variables that belong either to module $A$ or module $B$, and $\Psi_{(A\cup B)^\mathsf{c}}$ be the set of variables that do not belong to either module $A$ or $B$. Clearly $\Psi = \Psi_{A\setminus B}\, \cup\, \Psi_{B\setminus A}\, \cup\, \Psi_{A\cap B}\, \cup\, \Psi_{(A\cup B)^\mathsf{c}}$, as illustrated by the Venn diagram in Figure~\ref{F2}. We also define analogous partitions of $X$ and $\Theta$.

The following Lemma, illustrated in Figure~\ref{F2}, describes several properties of these sets, and the edges in the DAG between these sets, that will be used to derive the posterior and cut distributions.

\begin{lemma}
\label{l1}
Given a DAG $G=(\Psi,\mathcal{E})$ with modules $\Psi_A = (X_A,\Theta_A)$ and $\Psi_B = (X_B,\Theta_B)$ formed according to Rule~\ref{Ar1}, then the following statements about the partition $\Psi = \Psi_{A\setminus B}\, \cup\, \Psi_{B\setminus A}\, \cup\, \Psi_{A\cap B}\, \cup\, \Psi_{(A\cup B)^\mathsf{c}}$ hold.
\begin{enumerate}

    \item
        $\Psi_{(A\cup B)^\mathsf{c}}$ contains only parameters and has no children:
        \begin{itemize}
        \item $\Psi_{(A\cup B)^\mathsf{c}} = \Theta_{(A\cup B)^\mathsf{c}}$
        \item $\text{ch}(\Psi_{(A\cup B)^\mathsf{c}})=\emptyset$
        \end{itemize}
    \item
    The parents and children of $\Psi_{A\setminus B}$ are such that:
    \begin{itemize}
        \item $\text{pa}(\Psi_{A\setminus B}) \subseteq \Psi_{A\cap B}$
        \item $\text{ch}(\Psi_{A\setminus B}) \subseteq \Psi_{A\cap B} \cup \Psi_{(A\cup B)^\mathsf{c}}$
    \end{itemize}
    For $\Psi_{B\setminus A}$ the equivalent results hold.

    \item No v-structures $\psi_{A\setminus B}\rightarrow \psi_{A\cap B} \leftarrow \psi_{B\setminus A}$ exist with $\psi_{A\cap B}\in \Psi_{A\cap B}$, $\psi_{A\setminus B}\in \Psi_{A\setminus B}$ and $\psi_{B\setminus A}\in \Psi_{B\setminus A}$.

\item
    The parents of $\Theta_{A\cap B}$ and $X_{A\cap B}$ are such that:
    \begin{itemize}
    \item $\text{pa}(\Theta_{A\cap B}) \subseteq X_{A\cap B}$
    \item $\text{pa}(X_{A\cap B}) \subseteq \{\Theta_{A\setminus B}, \Theta_{A\cap B}, \Theta_{B\setminus A}, X_{A\setminus B},X_{B\setminus A}\}$
    \end{itemize}
\end{enumerate}
\end{lemma}

\subsubsection{Standard Bayesian inference}
We first consider the posterior distribution for $\Theta$ given $X$ under standard Bayesian inference. Note that the complete variable set $\Psi$ is naturally a self-contained Bayesian module.
\begin{lemma}
\label{l2}
Given a DAG $G=(\Psi,\mathcal{E})$ with modules $\Psi_A = (X_A,\Theta_A)$ and $\Psi_B = (X_B,\Theta_B)$ formed according to Rule~\ref{Ar1}, the standard posterior distribution for $\Theta$ can be written as either
\begin{equation*}
p(\Theta\given X)= p(\Theta_{(A\cup B)^\mathsf{c}}\given \text{pa}(\Theta_{(A\cup B)^\mathsf{c}})) \, p(\Theta_{B\setminus A}\given \Theta_{A\cap B},X_{B\setminus A},X_{A\cap B}) \, p(\Theta_A\given X)
\end{equation*}
or
\begin{equation*}
p(\Theta\given X)=p(\Theta_{(A\cup B)^\mathsf{c}}\given \text{pa}(\Theta_{(A\cup B)^\mathsf{c}})) \, p(\Theta_{A\setminus B}\given \Theta_{A\cap B},X_{A\setminus B},X_{A\cap B}) \, p(\Theta_B\given X).
\end{equation*}
\end{lemma}

This representation of the standard posterior distribution tells us that the inference of $\Theta_{A}$ might be affected by $\Theta_{B}$ (and vice versa) under the standard Bayesian framework.

\addtocounter{example}{-1}
\begin{example}[continued]
Consider the running example in Figure~\ref{F1}.
Suppose we partition the observables into the main study data $X_A^\ast = \{Y_{1:m}, Z_{1:m}, C_{1:m}\}$ and the external validation study data $X_B^\ast = \{Z_{(m+1):n}, W_{(m+1):n}, C_{(m+1):n}\}$.
By Rule~\ref{Ar1}, module $A$ is formed of the observables $X_A = X_A^\ast$ and the parameters $\Theta_A = \{\beta, W_{1:m}, \pi, \lambda\}$; and module $B$ is formed of the observables $X_B = X_B^\ast$ and the parameters $\Theta_B = \{\pi, \lambda\}$.
By the second formula in Lemma~\ref{l2}, the standard posterior distribution for $\Theta = (\Theta_A, \Theta_B)$ can be written as
\begin{equation*}
p(\beta, W_{1:m}, \pi, \lambda \given Y_{1:n}, Z_{1:n}, C_{1:n})
=
p(\beta, W_{1:m} \given \pi, \lambda, Y_{1:m}, Z_{1:m}, C_{1:m})\,
p(\pi, \lambda \given Y_{1:m}, Z_{1:n}, W_{(m+1):n}, C_{1:n}),
\end{equation*}
since $\Theta_{(A \cup B)^\mathsf{c}} = \emptyset$, $\Theta_{A \setminus B} = \{\beta, W_{1:m}\}$, $\Theta_{A \cap B} = \{\pi, \lambda\}$, $X_{A \setminus B} = X_A$ and $X_{A \cap B} = \emptyset$.
\end{example}

\subsubsection{Cut distributions}

Formulating cut distributions depends on understanding whether we can infer each module separately, without using the other module. This is equivalent to requiring both modules to be self-contained Bayesian modules, which the following theorem proves will be the case for modules formed using Rule~\ref{Ar1}.
\begin{theorem}
\label{l3}
For a DAG $G=(\Psi,\mathcal{E})$, the modules $\Psi_A = (X_A,\Theta_A)$ and $\Psi_B = (X_B,\Theta_B)$ formed according to Rule~\ref{Ar1} are both minimally self-contained Bayesian modules, with posterior distributions:
\begin{multline*}
p(\Theta_A\given X_A)\propto p(\Theta_{A\setminus B}\given \text{pa}(\Theta_{A\setminus B}))\,p(\Theta_{A\cap B}\given \text{pa}(\Theta_{A\cap B})) \\
\times p(X_{A\setminus B}\given \text{pa}(X_{A\setminus B}))\,p(X_{A\cap B}\cap X_A^\ast\given \text{pa}(X_{A\cap B}\cap X_A^\ast));
\end{multline*}
and
\begin{multline*}
p(\Theta_B\given X_B)\propto p(\Theta_{B\setminus A}\given \text{pa}(\Theta_{B\setminus A}))\,p(\Theta_{A\cap B}\given \text{pa}(\Theta_{A\cap B})) \\
\times p(X_{B\setminus A}\given \text{pa}(X_{B\setminus A}))\,p(X_{A\cap B}\cap X_B^\ast\given \text{pa}(X_{A\cap B}\cap X_B^\ast)).
\end{multline*}
\end{theorem}

\addtocounter{example}{-1}
\begin{example}[continued]
By Theorem~\ref{l3}, both modules $A$ and $B$ (defined above) are self-contained. The posterior distribution for module $A$ is
\begin{multline*}
p(\beta, W_{1:m}, \pi, \lambda \given Y_{1:m}, Z_{1:m}, C_{1:m})\\
\propto
\left\{p(\beta)\,
p(W_{1:m} \given \pi, C_{1:m})\right\}
\left\{p(\pi, \lambda)\right\}
\left\{p(Y_{1:m}, Z_{1:m}, C_{1:m} \given \beta, \lambda, W_{1:m})\right\},
\end{multline*}
where the braces group together terms from Theorem~\ref{l3}, and where the final factor is $\prod_{i=1}^{m}
p(Y_i \given \beta, W_i, C_i)\,
p(Z_i \given \lambda, W_i)$.
For module $B$ the posterior distribution is
\begin{equation*}
p(\pi, \lambda \given Z_{(m+1):n}, W_{(m+1):n}, C_{(m+1):n})\propto
\left\{p(\pi, \lambda)\right\}
\left\{p(Z_{(m+1):n}, W_{(m+1):n}, C_{(m+1):n} \given \lambda, \pi)\right\},
\end{equation*}
where the final factor is $\prod_{i=m+1}^{n}
p(Z_i \given \lambda, W_i)\,
p(W_i \given \pi, C_i)$.
\end{example}

To determine which influences of one module on the other module should be ``cut'', we need to identify the relationship between the modules. The second rule determines this.
\begin{RULE}[Identifying module ordering: two module case]
\label{Ar2}
For a DAG $G=(\Psi,\mathcal{E})$ with corresponding modules $A$ and $B$ formed by Rule~\ref{Ar1}, if a directed edge $\psi_1\rightarrow \psi_2$, with $\psi_1\in\Psi_{A\cap B}$ and $\psi_2\in\Psi_{B\setminus A}$, exists then denote this as $A\rightharpoonup B$ (or $B\leftharpoonup A$).
\begin{itemize}
\item If $A\rightharpoonup B$ holds but not $B\rightharpoonup A$, then module $A$ is the parent module and $B$ is the child module.
\item If both $A\rightharpoonup B$ and $B\rightharpoonup A$ holds, then either module can be parent and child module.
\item If neither $A\rightharpoonup B$ nor $B\rightharpoonup A$ hold, then the modules are unordered, which we denote by $(A,B)$.
\end{itemize}
\end{RULE}
When exactly one of $A\rightharpoonup B$ or $B\rightharpoonup A$ holds, then this order is fixed. However it is possible for both $A\rightharpoonup B$ and $B\rightharpoonup A$ to hold. In this case, the user is free to choose the order, but normally the child module should be chosen to be the more `suspect' module to prevent this module from affecting inference for the parent module.

The structure of the modules, and the role of module ordering, can be further understood by the following conditional independence result.
In particular, this Lemma implies that when modules are unordered, inference is completely separate between the modules even under standard Bayesian inference.

\begin{lemma}
\label{Th1}
Given a DAG $G=(\Psi,\mathcal{E})$ with modules $\Psi_A = (X_A,\Theta_A)$ and $\Psi_B = (X_B,\Theta_B)$ formed according to Rule~\ref{Ar1}, then $\Psi_{A\setminus B}$ and $\Psi_{B\setminus A}$ are d-separated by $\Psi_{A\cap B}$ in DAG $G$ (see the definition of d-separation in Appendix D in the supplementary materials) and we have
\[
\Psi_{A\setminus B}\indep \Psi_{B\setminus A} \given  \Psi_{A\cap B}.
\]
Furthermore, when the modules are unordered according to Rule 2, $\Psi_{A\cap B}=\emptyset$ and so $\Psi_{A\setminus B}\indep \Psi_{B\setminus A}$.
\end{lemma}

Now we consider cutting the influence of one of the modules. \thmcrossref{Theorem \ref{l3}} ensures that we can infer one module without being affected by the other module, but this inference depends on variables that are shared with the other module, such as $X_{A\cap B}\cap X_B^\ast$. Therefore, a viable choice of distribution for $\Theta$ given $X$ involves replacing $p(\Theta_A\given X)$ with $p(\Theta_A\given X_A)$ in the standard posterior distribution in \thmcrossref{Lemma \ref{l2}}. This is described in Rule~\ref{Ar3}, which involves a sub-graph of the DAG $G$ called a cut-sub-graph, which we first define.

\begin{definition}[$\Psi_0$-cut-sub-graph]
Given a DAG $G=(\Psi,\mathcal{E})$ and a set of variables $\Psi_0\subseteq\Psi$, we define the cut-sub-graph $G_{\text{cut}(\Psi_0)}=(\Psi,\mathcal{E}_0)$ in which edges from $\Psi\setminus\Psi_0$ to $\Psi_0$ are removed
\[
\mathcal{E}_0 = \left\{a\rightarrow b \in \mathcal{E}\ \given (b\in\Psi\setminus\Psi_0) \text{ or } (b\in\Psi_0) \cap (a\in\Psi_0)\right\}.
\]
\end{definition}

\begin{RULE}[Cutting feedback to the parent module]
\label{Ar3}
For a DAG $G=(\Psi,\mathcal{E})$ with modules $\Psi_A = (X_A,\Theta_A)$ and $\Psi_B = (X_B,\Theta_B)$ formed according to Rule~\ref{Ar1}, to cut the feedback from child module $B$ to parent module $A$, we prune the original DAG $G$ to obtain the $\Psi_{(B\setminus A)\cup (A\cup B)^\mathsf{c}}$-cut-sub-graph $G_{\text{cut}(\Psi_{(B\setminus A)\cup (A\cup B)^\mathsf{c}})}$. The component of the cut distribution for parameters $\Theta_A$ is the posterior distribution of $\Theta_A$ in $G_{\text{cut}(\Psi_{(B\setminus A)\cup (A\cup B)^\mathsf{c}})}$. Note that under $G_{\text{cut}(\Psi_{(B\setminus A)\cup (A\cup B)^\mathsf{c}})}$, this posterior distribution is $p(\Theta_A\given X_A)$.
\end{RULE}

Rule \ref{Ar3} tells us how to infer the parent module A, but does not describe how to infer the remaining parameters $\Psi_{B\setminus A}$ of the child module $B$. Note that, although module $B$ is self-contained, the parameters $\Theta_{A\cap B}$ that are in module B but also in module A have now been inferred according to Rule~\ref{Ar3}. Therefore, standard Bayesian inference cannot be immediately used within module $B$. We introduce the following definition.

\begin{definition}[Conditional self-contained Bayesian inference]
\label{DE2}
Consider a DAG $G=(\Psi,\mathcal{E})$ with two minimally self-contained Bayesian modules $\Psi_A=(X_A,\Theta_A)$ and $\Psi_B=(X_B,\Theta_B)$ formed according to Rule~\ref{Ar1}. Suppose that we choose to estimate a set of parameters $\Theta_{\mathrm{fix}}\subseteq\Theta_B$ using only module A or using only its own prior distribution $p(\Theta_{\mathrm{fix}}\given \text{pa}(\Theta_{\mathrm{fix}}))$. Then we say conditional self-contained Bayesian inference for $\Theta_{\mathrm{unfix}}=\Theta_B\setminus\Theta_{\mathrm{fix}}$ is conducted when the posterior distribution is given by
\[
p(\Theta_{\mathrm{unfix}}\given \Theta_{\mathrm{fix}},X_B)
\]
where for any $\theta^\ast\in\Theta_{\mathrm{fix}}$ we have either
\[
\theta^\ast\sim p(\theta^\ast\given \text{pa}(\theta^\ast))
\]
or
\[
\theta^\ast\sim \int p(\Theta_A\given X_A)\  d(\Theta_A\setminus\theta^\ast).
\]
\end{definition}

The aim of self-contained Bayesian modules is to prevent using information that is external to the module, whereas the aim of conditional self-contained Bayesian inference is to prevent using (at least some) information that is internal to a module. More specifically, it prevents inference for some parameters being affected by observable random variables within the same module.

We now return to the question of inference for module B under the two-module case when part of its parameters $\Theta_{A\cap B}$ have been inferred by its parent module $A$. To infer the remaining component $\Theta_{B\setminus A}$ we propose to conduct the conditional self-contained Bayesian inference. This lets us build the final piece of the cut distribution. We have the following rule:

\begin{RULE}[Cut distribution: two module case]
\label{Ar4}
Given a DAG $G=(\Psi,\mathcal{E})$ with modules $\Psi_A = (X_A,\Theta_A)$ and $\Psi_B = (X_B,\Theta_B)$ formed according to Rule~\ref{Ar1}, when the module order according to Rule~\ref{Ar2} is $A\rightharpoonup B$, the cut distribution for parameters $\Theta$ is:
\begin{equation*}
 p_{A\rightharpoonup B}(\Theta\given X):= p(\Theta_{(A\cup B)^\mathsf{c}}\given \text{pa}(\Theta_{(A\cup B)^\mathsf{c}})) \, p(\Theta_{B\setminus A}\given \Theta_{A\cap B},X_{B\setminus A},X_{A\cap B}) \, p(\Theta_A\given X_A).
\end{equation*}
Similarly, when the modules' order is $B\rightharpoonup A$, the cut distribution is
\begin{equation*}
p_{B\rightharpoonup A}(\Theta\given X):= p(\Theta_{(A\cup B)^\mathsf{c}}\given \text{pa}(\Theta_{(A\cup B)^\mathsf{c}}))
\, p(\Theta_{A\setminus B}\given \Theta_{A\cap B},X_{A\setminus B},X_{A\cap B}) \, p(\Theta_B\given X_B).
\end{equation*}
When two modules are unordered, the cut distribution is
\[
    p_{(A,B)}(\Theta\given X):= p(\Theta_{(A\cup B)^\mathsf{c}}\given \text{pa}(\Theta_{(A\cup B)^\mathsf{c}}))\,p(\Theta_B\given X_B)\,p(\Theta_A\given X_A).
\]
\end{RULE}
It is easy to check that both $p_{A\rightharpoonup B}$ and $p_{B\rightharpoonup A}$ are valid probability distributions but they might be different from either each other or the standard posterior distribution. When the module order is $A\rightharpoonup B$, the inference of parameters $\Theta_A$ is solely determined by the observable random variables $X_A$ from the same module. Hence, we conclude that under cut distribution $p_{A\rightharpoonup B}$, inference of module A is not influenced by information from $\Psi_{B\setminus A}$. This is in contrast to standard Bayesian inference or inference when assuming $B\rightharpoonup A$, in which inference of $\Theta_A$ might be influenced by variables in module $B$. Note that, the distribution of $\Theta_{(A\cup B)^\mathsf{c}}$ conditional on its parents $\text{pa}(\Theta_{(A\cup B)^\mathsf{c}})$ is unchanged no matter what kind of inference model we use and there is no feedback from $\Psi_{(A\cup B)^\mathsf{c}}$. In addition, $p_{(A,B)}(\Theta\given X)=p(\Theta\given X)$ when $\Psi_{A\cap B} = \emptyset$, which corresponds to modules $A$ and $B$ being unordered. Hence, we conclude that cut distribution $p_{A\rightharpoonup B}$ prevents information from variables $\Psi_{B\setminus A}$.

\addtocounter{example}{-1}
\begin{example}[continued]
For the same modules defined above, note that there are edges from $\Psi_{A \cap B} = \{\pi, \lambda\}$ both to $\Psi_{B \setminus A} = X_B$ and to $\Psi_{A \setminus B} = \Theta_{A \setminus B} \cup X_{A \setminus B}$.
This means that both $A\rightharpoonup B$ and $B\rightharpoonup A$ hold by Rule~\ref{Ar2}, and so we are free to choose the module ordering.
Since we regard module B that comes from an external validation study as more reliable than module $A$, the ordering $B\rightharpoonup A$ is appropriate.
Applying Rules~\ref{Ar3} and~\ref{Ar4}, the corresponding cut distribution is
\begin{multline*}
p_{B\rightharpoonup A}(\beta, W_{1:m} \given \pi, \lambda, Y_{1:m}, Z_{1:m}, C_{1:m})
=
p(\pi, \lambda \given Z_{(m+1):n}, W_{(m+1):n}, C_{(m+1):n})\\
\times p(\beta, W_{1:m} \given \pi, \lambda, Y_{1:m}, Z_{1:m}, C_{1:m})
\end{multline*}
Alternatively, if we choose the module ordering $A\rightharpoonup B$, then the cut distribution is
\begin{equation*}
 p_{A\rightharpoonup B}(\beta, W_{1:m} \given \pi, \lambda, Y_{1:m}, Z_{1:m}, C_{1:m})
=
p(\beta, W_{1:m}, \pi, \lambda \given Y_{1:m}, Z_{1:m}, C_{1:m})
\end{equation*}
but note that this distribution is unlikely to be suitable in applied settings, since it entirely disregards the data in module B because all parameters are in module A.
\end{example}

\subsubsection{Theoretical properties}
We now justify the cut distribution. When the module order is $A\rightharpoonup B$, we know that $p(\Theta_{(A\cup B)^\mathsf{c}}\given \text{pa}(\Theta_{(A\cup B)^\mathsf{c}}))$ is unchanged between the standard posterior distribution and cut distribution. The component $p(\Theta_A\given X_A)$ is justified by \thmcrossref{Theorem \ref{l3}}. Thus the only component that we need to justify is $p(\Theta_{B\setminus A}\given \Theta_{A\cap B}, X_{B\setminus A}, X_{A\cap B})$, which we will do in terms of the Kullback-Leibler (KL) divergence, defined for two arbitrary densities $f_1(x)$ and $f_2(x)$ as:
\[
\mathbb{D}_{KL}\left(f_1(\cdot), f_2(\cdot)\right) :=\int f_1(x) \log\left(f_1(x)/f_2(x)\right)dx.
\]
We have the following theorem:
\begin{theorem}
\label{Th2}
For a DAG $G=(\Psi,\mathcal{E})$ with modules $\Psi_A = (X_A,\Theta_A)$ and $\Psi_B = (X_B,\Theta_B)$ formed according to Rule~\ref{Ar1}, let $f(\Theta_{B\setminus A})$ be a probability density function for parameters $\Theta_{B\setminus A}$. Given the joint distribution $p(X,\Theta)$ and denote:
\[
p_f(\Theta)=p(\Theta_{(A\cup B)^\mathsf{c}}\given \text{pa}(\Theta_{(A\cup B)^\mathsf{c}}))\,f(\Theta_{B\setminus A})\, p(\Theta_A\given X_A),
\]
we have
\[
p(\Theta_{B\setminus A}\given \Theta_{A\cap B},X_{B\setminus A},X_{A\cap B})  \propto \argmin_{f(\Theta_{B\setminus A})} \mathbb{D}_{KL}\left(p_f(\cdot),p(X,\cdot)\right).
\]
\end{theorem}
This extends to our cut formulation the results of \textcite{yu2021variational} and \textcite{carmonaScalableSemiModularInference2022} that the standard two-module cut formulation and semi-modular posteriors, respectively, are the minimisers of KL divergences.

\subsection{Within-module cut}
When we regard some prior knowledge as more reliable than the information from the observations for some specific parameters, but inference for other parameters still depends on the observations, we can use a within-module manipulation. The within-module cut aims to protect the inference of some parameters from being affected by the suspect information from observations.
This requires a specific type of conditional self-contained Bayesian inference.
\textcite{carmonaScalableSemiModularInference2022} call this modulating prior feedback.

\addtocounter{example}{-1}
\begin{example}[continued]
Consider again the dashed module within the running example in Figure~\ref{F1}. Suppose we wish to estimate the true exposure $W_{1:m}$ without using any information from the outcomes $Y_{1:m}$. The posterior distribution for $W_{1:m}$ is clearly affected by the observable variable $Y_{1:m}$. An alternative distribution for $W_{1:m}$ is simply its marginal prior distribution $\int p(\pi) \prod_{i=1}^{m} p(W_{i} \given \pi, C_{i})d\pi$ (that conditions upon $C_{i}$ since the whole model is specified conditional upon these covariates).
\begin{equation*}
p_{\text{within}}(\beta, W_{1:m}) =
p(\beta \given Y_{1:m}, W_{1:m}, C_{1:m})
\int p(\pi) \prod_{i=1}^{m} p(W_{i} \given \pi,  C_{i})d\pi
\end{equation*}
This alternative distribution uses a within-module cut to ensure that our inference for $W_{1:m}$ uses no information from the associated observable variables $Y_{1:m}$.
\end{example}

\subsection{Three module case}\label{SE2.5}
Our extension to the case of split models into three modules involves two steps.
First, we split the model into two modules, as in Section~\ref{Sec2}.
We then form the third module by splitting either the child module or the parent module.
This sequential approach will form the basis for our approach to the general case of more than three modules in Section \ref{SE2.6}.
In this section, we discuss how three modules are constructed; how to determine the order of the three modules; and finally the corresponding cut distributions.

\subsubsection{Module construction}
Each of the three modules we construct corresponds to a component of a partition of the observable random variables $X$ into three sets $X_A^\ast \cup X_B^\ast \cup X_C^\ast$.
Each module is formed from each component of this partition using Rule~\ref{Ar1}.
However, to consider module ordering and to facilitate extension to more than three modules, we consider these three modules as arising from splitting one of the modules in a two module split.

Specifically, let $X_T^\ast = X_B^\ast \cup X_C^\ast$, and denote by $A$ and $T$ the two modules formed by Rule~\ref{Ar1} from the partition $X = X_A^\ast \cup X_T^\ast$ of the observable random variables.
Recall that some parameters may belong to $(A\cup T)^\mathsf{c}$ and so are not part of either module.
To obtain the third module we split module $T$. Module $T$ originates from observable random variables $X_T^\ast$, so we split this set of observable random variables back into the two disjoint sets $X_B^\ast$ and $X_C^\ast$. We then identify module $B$ using Rule~\ref{Ar1} for $X_B^\ast$; and module $C$ by applying Rule~\ref{Ar1} for $X_C^\ast$. In this way, we obtain three modules $A$, $B$ and $C$, where $\Psi_B\cup \Psi_C=\Psi_T$.
\begin{figure}[t]
\setlength{\abovecaptionskip}{0cm}
\setlength{\belowcaptionskip}{0cm}
\centering
\includegraphics[width=0.5\textwidth]{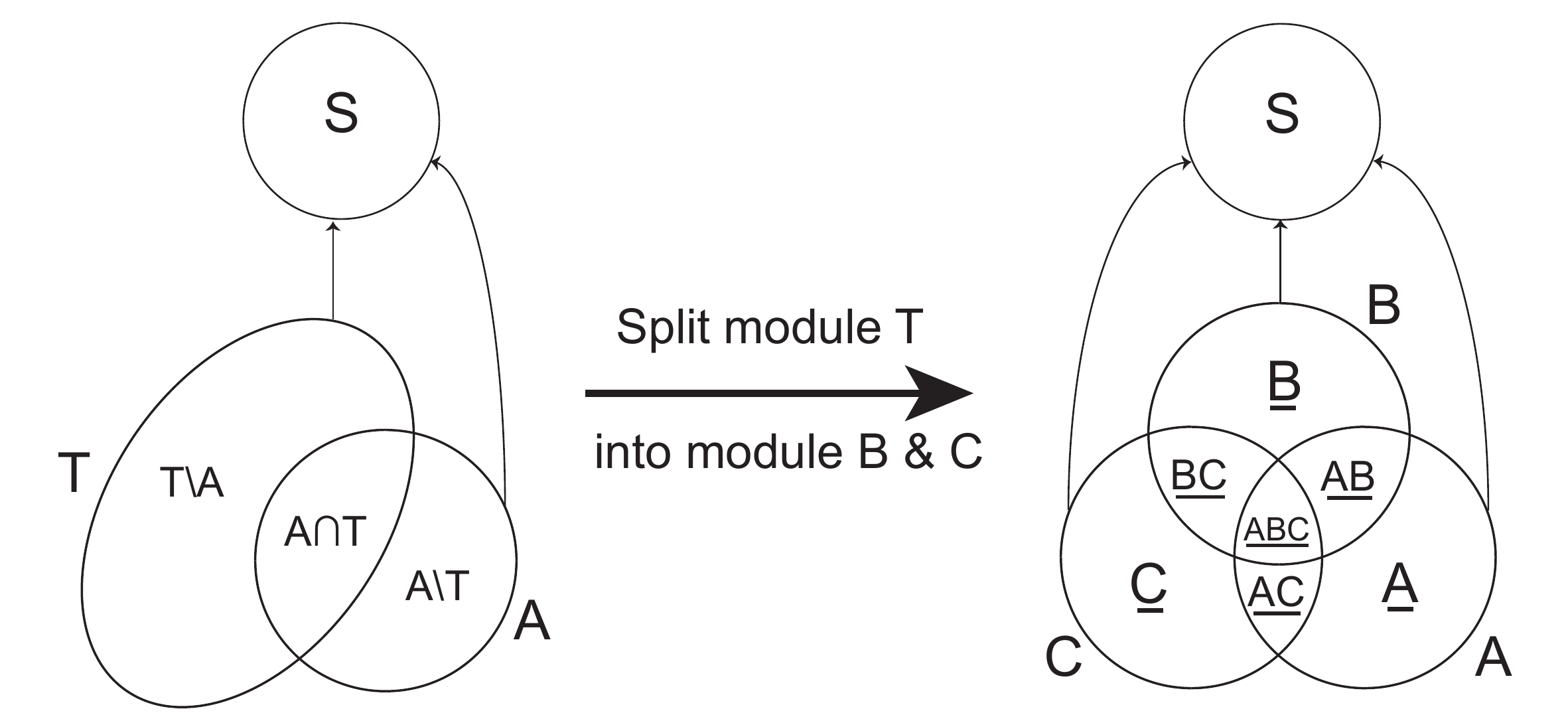}
\caption[Venn diagram illustrating the splitting of module $T$ into modules $B$ and $C$.]{\textbf{Venn diagram illustrating the splitting of module $T$ into modules $B$ and $C$.} The arrows represent all directed edges from modules to $S=(A\cup T)^\mathsf{c}$.}
\label{F4}
\end{figure}

\subsubsection{Module ordering}
To identify the ordering of modules $A$, $B$ and $C$, first note that prior to splitting module $T$ we can determine the ordering of modules $A$ and $T$ using Rule \ref{Ar2}.
To identify the order after splitting module $T$, we consider a partition of the variables $\Psi_{A\cup B\cup C}$ into 7 disjoint groups: $\Psi_{\underline{A}}$, $\Psi_{\underline{AB}}$, $\Psi_{\underline{AC}}$, $\Psi_{\underline{ABC}}$, $\Psi_{\underline{B}}$, $\Psi_{\underline{BC}}$ and $\Psi_{\underline{C}}$, as shown in Figure \ref{F4}.
\begin{align*}
\underline{A} &= A\setminus \left(B\cup C\right) \ &\underline{B} &= B\setminus \left(A\cup C\right) \ &\underline{C} &= C\setminus \left(A\cup B\right) \\
\underline{AB} &= \left(A\cap B\right)\setminus C \ &\underline{AC} &= \left(A\cap C\right)\setminus B\ &\underline{BC} &= \left(B\cap C\right)\setminus A \\
&& \underline{ABC} &= A\cap B \cap C
\end{align*}
We also define $S := (A\cup \cup B \cup C)^\mathsf{c}$.
The following lemma extends Lemma~\ref{Th1} to the three module setting:
\begin{lemma}
\label{l4}
For a DAG $G=(\Psi,\mathcal{E})$ with corresponding modules $A$, $B$ and $C$ and $T$, with $\Psi_T=\Psi_B\cup \Psi_C$, if $\Psi_{\underline{BC}}\neq\emptyset$, then there must be at least one edge $a\rightarrow b$ where $a\in\Psi_{\underline{BC}}$ and $b\in(\Psi_{\underline{B}}\cup X_B^\ast)\cup(\Psi_{\underline{C}}\cup X_C^\ast)$. Otherwise if $\Psi_{\underline{BC}}= \emptyset$, we have
\[
\Psi_{\underline{B}}\indep \Psi_{\underline{C}} \given  \Psi_{A}.
\]
\end{lemma}

Given \thmcrossref{Lemma \ref{l4}}, we give the following Rule to determine the ordering of the three modules.
\begin{RULE}[Identifying module ordering: three module case]
\label{Ar5}
For a DAG $G=(\Psi,\mathcal{E})$ with corresponding modules $A$, $B$ and $C$ and $T$, with $\Psi_T=\Psi_B\cup \Psi_C$, there are three possible scenarios, depending on the ordering of modules $A$ and $T$ (determined by Rule~\ref{Ar2}):
\begin{enumerate}
    \item When the order prior to splitting is $A\rightharpoonup (B\cup C)$, we will split the child module. In this case, by Rule~\ref{Ar2}, at least one of $\Psi_{A \cap B}$ and $\Psi_{A \cap C}$ must be non-empty.
    \begin{enumerate}
    \item When both $\Psi_{A\cap B}\neq\emptyset$ and $\Psi_{A\cap C}\neq\emptyset$.
        \begin{enumerate}
            \item If there is at least one edge $b\rightarrow c$ where $b\in\Psi_{\underline{BC}}$ and $c\in(\Psi_{\underline{C}}\cup X_C^\ast)$, we say module $C$ is the child module of module $B$ and module $B$ is the child module of module $A$. We denote this as $A\rightharpoonup B\rightharpoonup C$. Similarly if there is at least one edge $c\rightarrow b$ where $c\in\Psi_{\underline{BC}}$ and $b\in(\Psi_{\underline{B}}\cup X_B^\ast)$, we denote this as $A\rightharpoonup C\rightharpoonup B$ .
            \item If $\Psi_{\underline{BC}}=\emptyset$, we say module $B$ and $C$ are unordered and they are child modules of module $A$. We denote this as $A\rightharpoonup (B, C)$.
        \end{enumerate}

    \item When only $\Psi_{A \cap B}$ or $\Psi_{A \cap C}$ is not empty. Without loss of generality, we consider only the case when $\Psi_{A \cap C} = \emptyset$:
        \begin{enumerate}
            \item If there is at least one edge $b\rightarrow c$ where $b\in\Psi_{\underline{BC}}$ and $c\in(\Psi_{\underline{C}}\cup X_C^\ast)$, we say module $C$ is the child module of module $B$ and module $B$ is the child module of module $A$. We denote this as $A\rightharpoonup B\rightharpoonup C$.
            \item If there is at least one edge $c\rightarrow b$ where $c\in\Psi_{\underline{BC}}$ and $b\in(\Psi_{\underline{B}}\cup X_B^\ast)$, we say module $A$ and $C$ are unordered and they are parent modules of module $B$. We denote this as $(A, C)\rightharpoonup B$.
            \item  If $\Psi_{\underline{BC}}=\emptyset$, we say module $B$ is the child module of module $A$ and these modules are unordered with module $C$. We denote this as $(C,(A\rightharpoonup B))$.
        \end{enumerate}
    \end{enumerate}
    \item When the order prior to splitting is $(B\cup C)\rightharpoonup A$, we will split the parent module. In this case at least one of $\Psi_{A \cap B}$ and $\Psi_{A \cap C}$ must be non-empty.
    \begin{enumerate}
        \item When both $\Psi_{A\cap B}\neq\emptyset$ and $\Psi_{A\cap C}\neq\emptyset$.
        \begin{enumerate}
            \item If there is at least one edge $b\rightarrow c$ where $b\in \Psi_{B\cap C}$ and $c\in\Psi_{B\setminus C}$, we denote this as $B\rightharpoonup C\rightharpoonup A$. Similarly if there is at least one edge $c\rightarrow b$ where $c\in \Psi_{B\cap C}$ and $b\in\Psi_{C\setminus B}$ we denote this as $C\rightharpoonup B\rightharpoonup A$.
            \item  If $ \Psi_{B\cap C}=\emptyset$, we say module $B$ and $C$ are unordered, and we denote this as $(B, C)\rightharpoonup A$.
        \end{enumerate}
        \item When only $\Psi_{A \cap B}$ or $\Psi_{A \cap C}$ is not empty. Without loss of generality, we consider only the case when $\Psi_{A \cap C} = \emptyset$:
        \begin{enumerate}
            \item If there is at least one edge $b\rightarrow c$ where $b\in\Psi_{\underline{BC}}$ and $c\in(\Psi_{\underline{C}}\cup X_C^\ast)$, we say module $A$ and module $C$ are unordered, and these modules are child modules of module $B$. We denote this as $B\rightharpoonup (A, C)$.
            \item If there is at least one edge $c\rightarrow b$ where $c\in\Psi_{\underline{BC}}$ and $b\in(\Psi_{\underline{B}}\cup X_B^\ast)$, we denote this as $C\rightharpoonup B\rightharpoonup A$.
            \item If $\Psi_{\underline{BC}}=\emptyset$, we say module $A$ is the child module of module $B$ and these modules are unordered with module $C$. We denote this as $(C,(B\rightharpoonup A))$.
        \end{enumerate}

    \end{enumerate}
    \item When the order prior to splitting is $(A, (B\cup C))$, meaning that modules $A$ and $B\cup C$ are unordered, then both $\Psi_{A \cap B}$ and $\Psi_{A \cap C}$ must be empty.
    \begin{enumerate}
        \item If there is at least one edge $b\rightarrow c$ where $b\in\Psi_{\underline{BC}}$ and $c\in(\Psi_{\underline{C}}\cup X_C^\ast)$, we say module $C$ is the child module of module $B$ and these modules are unordered with module $A$. We denote this as $(A,(B\rightharpoonup C))$. Correspondingly we denote the ordering as $(A,(C\rightharpoonup B))$ if there is at least one edge $c\rightarrow b$ where $c\in\Psi_{\underline{BC}}$ and $b\in(\Psi_{\underline{B}}\cup X_B^\ast)$.
        \item If $\Psi_{\underline{BC}}=\emptyset$, we say modules $A$, $B$ and $C$ are unordered, and denote this as $(A,B,C)$.
    \end{enumerate}
\end{enumerate}
\end{RULE}
When considering three (or more) modules, it is helpful to represent the relationship between the modules graphically using a module ordering graph.

\begin{figure}[t]
\setlength{\abovecaptionskip}{0cm}
\setlength{\belowcaptionskip}{0cm}
\centering
\includegraphics[width=0.7\textwidth]{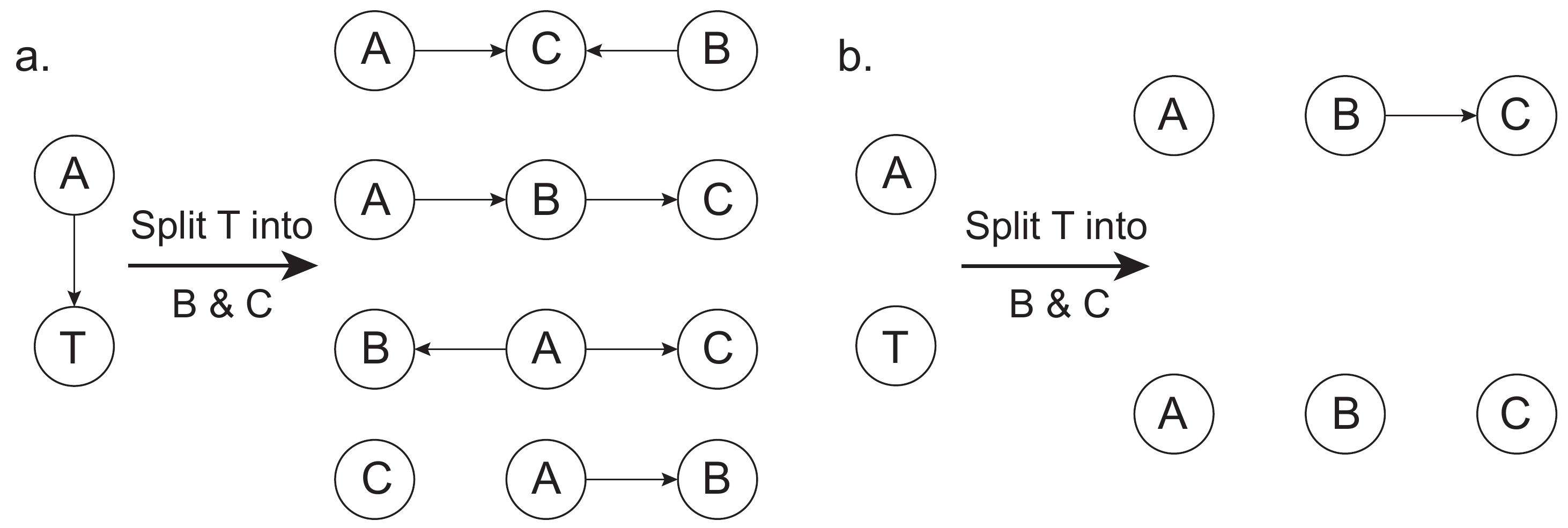}
\caption[Module order graphs when splitting two modules into three modules]{\textbf{Module order graphs when splitting two modules into three modules.} Circles represent modules and arrows reflect the direction of ordering between two modules. Each panel depicts the possible module orderings of three modules when splitting module $T$. Panel (a) shows the orderings when the original order is $A\rightharpoonup T$: from top to bottom $(A, B)\rightharpoonup C$; $A\rightharpoonup B\rightharpoonup C$; $A\rightharpoonup (B, C)$; and $(C, (A\rightharpoonup B))$. Panel (b) shows the orderings when the original two modules $A$ and $T$ are unordered: from top to bottom $(A, (B\rightharpoonup C))$; and $(A, B, C)$.}
\label{F5}
\end{figure}

\begin{definition}[Module ordering graph]
\label{DEmoduleorder}
For modules $M_1, \dots, M_s$, a module ordering graph $J$ is a DAG with nodes $M_1, \dots, M_s$, and with directed edges from $M_i$ to $M_j$ if module $M_i$ is a parent module of module $M_j$, for $i, j = 1, \dots, s$.
\end{definition}%
Figure \ref{F5} shows some example module ordering graphs when splitting two modules into three modules.

\subsubsection{Cut distributions}
We can now define the cut distribution for the three-module case in a similar manner to the two-module case.
We define the sets $\underline{\underline{A}} := A\setminus \underline{A}$, $\underline{\underline{B}} := B\setminus \underline{B}$, and $\underline{\underline{C}} := C\setminus \underline{C}$.
\begin{definition}[Cut distribution: three module case]
\label{DE3}
For a DAG $G=(\Psi,\mathcal{E})$ with corresponding modules $A$, $B$ and $C$, with module ordering determined by Rule~\ref{Ar5}, the corresponding cut distributions (where $A$, $B$ and $C$ are interchangeable) are:
\begin{enumerate}
    \item
$p_{A\rightharpoonup B\rightharpoonup C}(\Theta\given X) := p(\Theta_S\given \Theta_{A\cup B\cup C},X)\,p(\Theta_{\underline{C}}\given \Theta_{\underline{\underline{C}}},X_C)\,p(\Theta_{B\setminus A}\given \Theta_{A\cap B},X_B)\,p(\Theta_A\given X_A).$

    \item
$\begin{aligned}
p_{A\rightharpoonup (B, C)}(\Theta\given X) & := p(\Theta_S\given \Theta_{A\cup B\cup C},X)\,p(\Theta_{\underline{C}}\given \Theta_{A\cap C},X_C)\,p(\Theta_{\underline{B}}\given \Theta_{A\cap B},X_B)\,p(\Theta_A\given X_A).
\end{aligned}
$
    \item
$\begin{aligned}
 p_{(A, B)\rightharpoonup C}(\Theta\given X) &:= p(\Theta_S\given \Theta_{A\cup B\cup C},X)\,p(\Theta_{\underline{C}}\given \Theta_{\underline{\underline{C}}},X_C)\,p(\Theta_B\given X_B)\,p(\Theta_A\given X_A).
\end{aligned}
$
    \item
$\begin{aligned}
 p_{(A, (B\rightharpoonup C))}(\Theta\given X) &:= p(\Theta_S\given \Theta_{A\cup B\cup C},X)\,p(\Theta_{\underline{C}}\given \Theta_{B\cap C},X_C)\,p(\Theta_B\given X_B)\,p(\Theta_A\given X_A).
 \end{aligned}
$
    \item
$\begin{aligned}
 p_{(A, B, C)}(\Theta\given X) &:= p(\Theta_S\given \Theta_{A\cup B\cup C},X)\,p(\Theta_C\given X_C)\,p(\Theta_B\given X_B)\,p(\Theta_A\given X_A).
 \end{aligned}
$
\end{enumerate}
\end{definition}

Importantly, Definition \ref{DE3} indicates that, when splitting a two-module case to a three-module case, the component of the cut distribution for the module that is not split remains unchanged. This important property enables us to modify only the components of a cut distribution that involve modules being split while keeping all other components of the cut distribution unchanged.

We discuss the construction of the cut distribution by using the example of modules ordered as $A\rightharpoonup B\rightharpoonup C$. As in the two-module case, $p(\Theta_S\given \text{pa}(\Theta_S))=p(\Theta_S\given \Theta_{A\cup B\cup C},X)$ is directly borrowed from the joint distribution \eqref{E8} to form the component of the cut distribution for $\Theta_S$. Next, the procedure is the same as Rule \ref{Ar3}, except that conditional self-contained Bayesian inferences are conducted for module $B$ (conditional on everything that has been inferred in module $A$) and $C$ (conditional on everything that has been inferred in module $A$ and $B$). If we regard module $B$ and $C$ as a single module, then by Rule \ref{Ar3} we can write the component of the cut distribution for parameters $\theta\in\Theta_A$: $p(\Theta_A\given X_A)$, which is exactly the same as the component of the cut distribution in the two-module case (see \thmcrossref{Theorem \ref{l3}}). That is:
\begin{equation*}
p(\Theta_A\given X_A)\propto  p(\Theta_{A}\given \text{pa}(\Theta_{A}))\,p(X_{\underline{A}}\given \text{pa}(X_{\underline{A}}))\,p(X_{A\cap (B\cup C)}\cap X_A^\ast\given \text{pa}(X_{A\cap (B\cup C)}\cap X_A^\ast)).
\end{equation*}
The next set of parameters of interest involves parameters in module $B$ that are not in module A. These are $\Theta_{\underline{B}}$ and $\Theta_{\underline{BC}}$, since  $\Theta_{B\setminus A} = \Theta_{\underline{B}} \cup \Theta_{\underline{BC}}$. Conditional on information from module $A$, in order to prevent information from module $C$, we conduct conditional self-contained Bayesian inference for module $B$ by utilizing the following components of the joint distribution \eqref{E8}, which involve the distribution of $\Theta_{B\setminus A}$ and distributions conditioning on any variables from $\Theta_{B\setminus A}$ but neither conditioning on nor involving any variables from $\Psi_{\underline{C}}$, to form the component of the cut distribution of $\Theta_{B\setminus A}$:
\begin{multline*}
p(\Theta_{B\setminus A}\given \Theta_A,X_{A\cup B})
\propto p(\Theta_{\underline{B}}\given \text{pa}(\Theta_{\underline{B}}))\,p(\Theta_{\underline{BC}}\given \text{pa}(\Theta_{\underline{BC}}))\,p(X_{\underline{B}}\given \text{pa}(X_{\underline{B}}))  \\
 \times p(X_{\underline{AB}}\cap X_B^\ast\given \text{pa}(X_{\underline{AB}}\cap X_B^\ast))\,p(X_{\underline{BC}}\cap X_B^\ast\given \text{pa}(X_{\underline{BC}}\cap X_B^\ast)) \\
\times p(X_{\underline{ABC}}\cap X_B^\ast\given \text{pa}(X_{\underline{ABC}}\cap X_B^\ast)),
\end{multline*}
where the proportionality holds here because $\Theta_A$ has been inferred in module $A$. In addition, we have
\begin{align*}
\text{pa}(\Theta_{\underline{B}}) &\subseteq \{ \Theta_{\underline{\underline{B}}}, X_B \}\\ \text{pa}(\Theta_{\underline{BC}}) &\subseteq \{ \Theta_{\underline{ABC}},X_{B\cap C} \} \\
 \text{pa}(X_{\underline{B}}) &\subseteq \{\Theta_B, X_{\underline{\underline{B}}}\} \\
\text{pa}(X_{\underline{AB}}\cap X_B^\ast) &\subseteq \{\Theta_B, X_B\setminus X_{\underline{AB}}, X_{\underline{AB}}\cap X_A^\ast\}\\
\text{pa}(X_{\underline{BC}}\cap X_B^\ast) &\subseteq \{\Theta_B, X_B\setminus X_{\underline{BC}}, X_{\underline{BC}}\cap X_C^\ast\} \\
\text{pa}(X_{\underline{ABC}}\cap X_B^\ast) &\subseteq \left\{\Theta_B, X_B\setminus X_{\underline{ABC}},X_{\underline{ABC}}\cap (X_A^\ast\cup X_C^\ast)\right\},
\end{align*}
where all parameters involved in these sets of parent nodes belong to $\Theta_B=\Theta_{B\setminus A}\cup \Theta_{A\cap B}$ and $\Theta_{A\cap B}$ has been inferred in module $A$. Hence we can write $p(\Theta_{B\setminus A}\given \Theta_A,X_{A\cup B})=p(\Theta_{B\setminus A}\given \Theta_{A\cap B},X_B)$. Now we conduct conditional self-contained Bayesian inference for module $C$ by utilizing all remaining components of the joint distribution \eqref{E8} to form the final piece of component of the cut distribution for $\Theta_{\underline{C}}$:
\begin{multline*}
p(\Theta_{\underline{C}}\given \Theta_{A\cup B},X)\propto p(\Theta_{\underline{C}}\given \text{pa}(\Theta_{\underline{C}}))\, p(X_{\underline{C}}\given \text{pa}(X_{\underline{C}})) \, p(X_{\underline{BC}}\cap X_C^\ast\given \text{pa}(X_{\underline{BC}}\cap X_C^\ast)) \\
\times p(X_{\underline{AC}}\cap X_C^\ast\given \text{pa}(X_{\underline{AC}}\cap X_C^\ast)) \,p(X_{\underline{ABC}}\cap X_C^\ast\given \text{pa}(X_{\underline{ABC}}\cap X_C^\ast)),
\end{multline*}
where we similarly have
\begin{align*}
\text{pa}(\Theta_{\underline{C}}) &\subseteq \{ \Theta_{\underline{\underline{C}}}, X_C \}\\  \text{pa}(X_{\underline{C}}) &\subseteq \{\Theta_C, X_{\underline{\underline{C}}}\} \\
\text{pa}(X_{\underline{BC}}\cap X_C^\ast) &\subseteq \left\{\Theta_C, X_C\setminus X_{\underline{BC}}, X_{\underline{BC}}\cap X_B^\ast\right\}\\
\text{pa}(X_{\underline{AC}}\cap X_C^\ast) &\subseteq \left\{\Theta_C, X_C\setminus X_{\underline{AC}}, X_{\underline{AC}}\cap X_A^\ast\right\} \\
\text{pa}(X_{\underline{ABC}}\cap X_C^\ast) &\subseteq \left\{\Theta_C, X_C\setminus X_{\underline{ABC}},X_{\underline{ABC}}\cap (X_A^\ast\cup X_B^\ast)\right\}.
\end{align*}
Hence, we conclude that $p(\Theta_{\underline{C}}\given \Theta_{A\cup B},X)=p(\Theta_{\underline{C}}\given \Theta_{\underline{\underline{C}}},X_C)$.

In summary, the essential procedure is to split the observable random variables into three disjoint groups and enlarge these groups according to Rule \ref{Ar1}, then determine the ordering of the modules. The cut distribution in three module case is then formed by four components: a component each of three modules and a component for $\Theta_S$.

\addtocounter{example}{-1}
\begin{example}[continued]
Consider again the running example in Figure~\ref{F1}.
Suppose we partition the observables into $X_A^\ast = \{W_{(m+1):n}, C_{(m+1):n}\}$, $X_B^\ast = \{Z_{(m+1):n}\}$ and $X_C^\ast = \{Y_{1:m}, Z_{1:m}, C_{1:m}\}$.
By Rule~\ref{Ar1}, module $A$ is formed of the observables $X_A = X_A^\ast$ and the parameters $\Theta_A = \{\pi\}$; module $B$ is formed of the observables $X_B = X_B^\ast \cup \{W_{(m+1):n}\}$ and the parameters $\Theta_B = \{\lambda\}$; and module $C$ is formed of the observables $X_C = X_C^\ast$ and the parameters $\Theta_C = \{\beta, W_{1:m}, \pi, \lambda\}$.

Let us initially consider modules $B$ and $C$ together as module $T$.
Note that there are edges from $\Psi_{A \cap T} = \{\pi, W\}$ to $\Psi_{T \setminus A}$, but no edges from $\Psi_{A \cap T}$ to $\Psi_{A \setminus T} = \{C_{(m+1):n}\}$, so only $A \rightharpoonup T$ holds by Rule~\ref{Ar2}.

We now consider splitting module $T$ into modules $A$ and $B$ using Rule~\ref{Ar5}. We are splitting the child module, and both $\Psi_{A \cap B} \neq 0$ and $\Psi_{A \cap C} \neq 0$, so Rule~\ref{Ar5}(a) applies.
Now, $\Psi_{\underline{BC}} = \{\lambda\}$, and there exist edges from $\lambda$ to both $\Psi_{\underline{C}} \cup X_C^\ast = \{Y_{1:m}, Z_{1:m}, C_{1:m}, \beta, W_{1:m}\}$ and
$\Psi_{\underline{B}} \cup X_B^\ast = \{Z_{(m+1):n}\}$.
Thus by Rule~\ref{Ar5}(a)(i), both $A \rightharpoonup B\rightharpoonup C$ and $A \rightharpoonup C \rightharpoonup B$ hold, so are free to choose either module order.
Given that the external validation surrogate exposure data are regarded as more reliable than the main study data, the order $A \rightharpoonup B\rightharpoonup C$ may be preferred.
The corresponding cut distribution given by Definition~\ref{DE3} is
\begin{multline*}
p_{A\rightharpoonup B\rightharpoonup C}(\beta, W_{1:m}, \lambda, \pi \given Y_{1:m}, Z_{1:m}, C_{1:m}, Z_{(m+1):n}, W_{(m+1):n}, C_{(m+1):n})\\
= 
p(\beta, W_{1:m} \given \pi, \lambda, Y_{1:m}, Z_{1:m}, C_{1:m})\,
p(\lambda \given Z_{(m+1):n}, W_{(m+1):n})\,
p(\pi \given W_{(m+1):n}, C_{(m+1):n}),
\end{multline*}
noting that $\Psi_S = \emptyset$ and so the term for $\Theta_S$ is omitted.
\end{example}

\subsection{Multiple-module case: Sequential splitting technique} \label{SE2.6}
To construct our general modularisation framework involving an arbitrary number of modules, we continue to split modules sequentially. Specifically we consider splitting an arbitrary module $T$.
Modules may be related in a potentially complex structure, but we can reduce this complex structure of modules into the simple structure that we have considered already by grouping together the ancestor and descendant modules of $T$.
We will consider the case when the module to be split has both parent modules and child modules. The cases when the module has only parent modules or child modules are easy extensions.

\subsubsection{Module groups}
To simplify the complexity of the multiple module case, we group together modules into groups, and form a graphical representation of the relationship between these module groups.

\begin{definition}[Grouped module ordering graph]
\label{DEgroupmoduleordergraph}
For module groups $M_1^\ast, \dots, M_{s^\ast}^\ast$, where $M_i^\ast \subseteq \{M_1, \dots, M_s\}$ for all $i = 1, \dots, s^\ast$, and $M_i^\ast \cap M_j^\ast = \emptyset$, for any $i\neq j$, the grouped module ordering graph is a DAG with nodes $M_1^\ast$, \dots, $M_{s^\ast}^\ast$ and with directed edges from $M_i^\ast$ to $M_j^\ast$ if module group $M_i^\ast$ is an ancestor of $M_j^\ast$.
\end{definition}

The module groups we use are defined in terms of the ancestor and descendant modules of module $T$ in the module order graph $J$ prior to splitting module $T$.
\begin{definition}[Ancestor and descendant modules]
For a module $T$ within an ordering graph $J$
\begin{enumerate}
    \item If there is a direct path $A\rightsquigarrow T$ between two modules $A$ and $T$ with a path $A=M_1,M_2,\dots,M_s=T$ in the ordering graph $J$, then module $A$ is an ancestor module of module $T$.
    \item If there is a direct path $T\rightsquigarrow A$ between two modules $A$ and $T$ with a path $T=M_1,M_2,\dots,M_s=A$ in the ordering graph $J$, then module $A$ is a descendant module of module $T$.
\end{enumerate}
\end{definition}

\begin{figure}[!t]
\centering
\includegraphics[width=.7\textwidth]{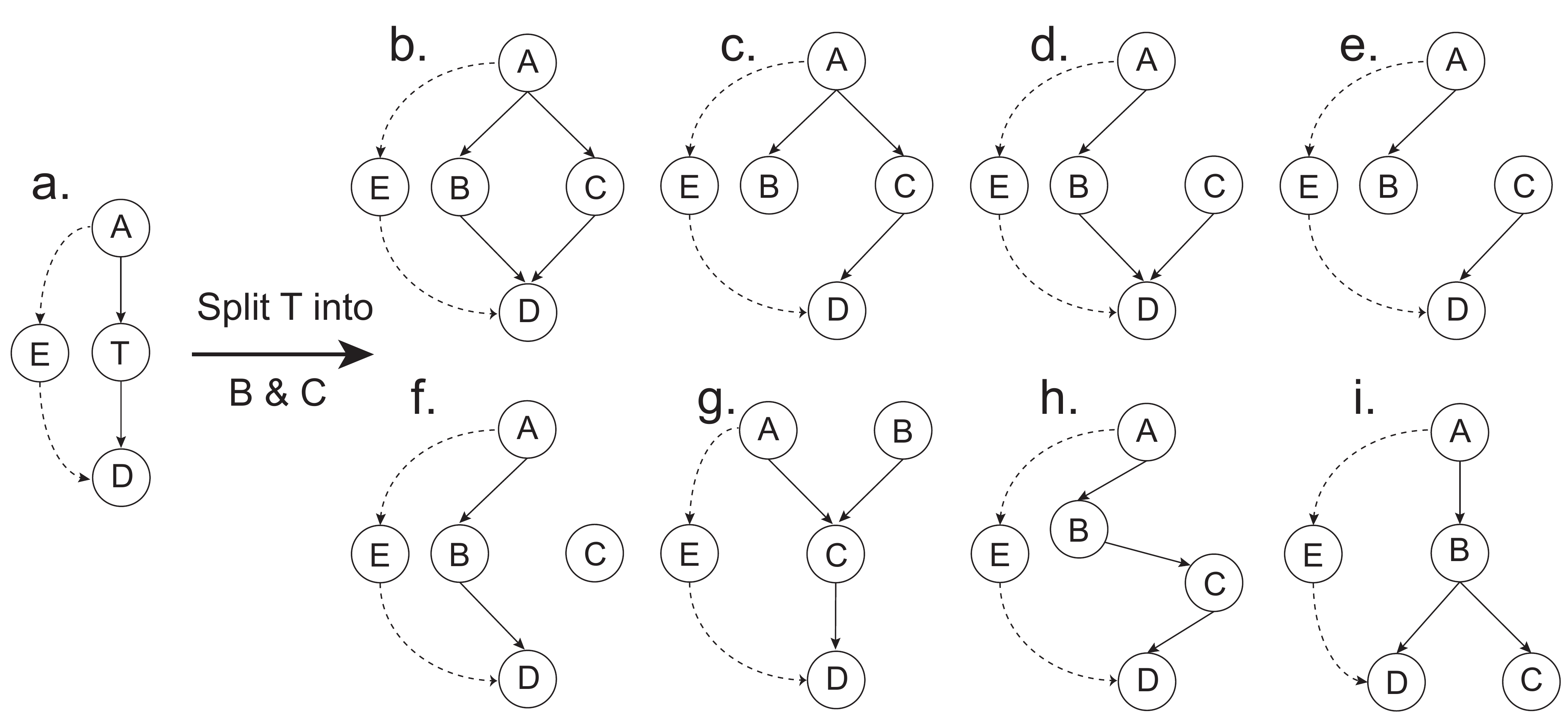}
\caption[Grouped module order graphs when splitting module $T$ with ancestor modules and descendant modules]{\textbf{Grouped module order graphs when splitting module $T$ with both ancestor modules $A$ and descendant modules $D$.} Circles represent modules or grouped modules. Solid arrows indicate the direction of the ordering between two modules or module groups. Dashed arrows represent a possible direction of ordering between two modules or module groups. (a) the original grouped module order graph before splitting module $T$. Panels (b)-(i) depict all possible grouped module order graphs after splitting module $T$ in modules $B$ and $C$.
}
\label{F7}
\end{figure}

The module groups we will use are the module $T$ itself, along with groups formed by the ancestors and descendants of $T$ in the ordering graph $J$, and a group containing all remaining modules.
Specifically, let $A$ denote the group of modules that are ancestor modules of $T$; let $D$ denote the group of modules that are descendant modules of $T$; and let $E$ denote the modules that are neither an ancestor module nor a descendant module of module $T$. Figure~\ref{F7}(a) shows the grouped module ordering graph for $A$, $T$, $D$ and $E$.

First consider the cut distribution prior to splitting module $T$. It is clear that inference of parameters $\Theta_A$ in module group $A$ is completely independent from all other modules, so the component of the cut distribution for module group $A$ is simply $p(\Theta_A\given X_A)$ since module $A$ is a self-contained Bayesian property. Conditional on module group $A$, inference of parameters within module groups $T$ and $E$ that are not inferred by module group $A$ is independent. Hence, conditional self-contained Bayesian inference is conducted and the components of the cut distribution for module groups $T$ and $E$ are $p(\Theta_{T\setminus A}\given \Theta_{A\cap T}, X_T)$ and $p(\Theta_{E\setminus A}\given \Theta_{A\cap E}, X_E)$. The final pieces of the cut distribution are the components for module group $D$ which is $p(\Theta_{D\setminus (A\cup T\cup E)}\given \Theta_{(A\cup T\cup E)\cap D}, X_D)$, and the component for $\Theta_{(A\cup T\cup D\cup E)^\mathsf{c}}$. Thus the cut distribution prior to splitting module $T$ is:
\begin{multline*}
p(\Theta\given X) = p(\Theta_{(A\cup T\cup D\cup E)^\mathsf{c}}\given \text{pa}(\Theta_{(A\cup T\cup D\cup E)^\mathsf{c}}))
\, p(\Theta_{D\setminus (A\cup T\cup E)}\given \Theta_{(A\cup T\cup E)\cap D}, X_D)\\
 \times p(\Theta_{E\setminus A}\given \Theta_{A\cap E}, X_E) \, p(\Theta_{T\setminus A}\given \Theta_{A\cap T}, X_T)  \, p(\Theta_A\given X_A)
\end{multline*}
This follows as a consequence of the following observations. We know that the component of the cut distribution for $\Theta_{(A\cup T\cup D\cup E)^\mathsf{c}}$, conditional on its parents, is always unchanged. If we regard module groups $A$, $T$ and $E$ as a single module, as in the two-module case, we know that the split of $A\cup T\cup E$ does not affect the component of module group $D$ in the cut distribution. Similarly, if we regard module groups $T$, $E$ and $D$ as a single module, as in the two-module case, the split of $T\cup E\cup D$ does not affect the component of module group $A$ in the cut distribution. Because inference of module group $E$ is independent from module $T$ conditional on module group $A$ and the inference of module group $A$ is unchanged, the split of module $T$ does not affect the inference of module group $E$.

\subsubsection{Updating the module order graph}
To determine the ordering of the modules after splitting $T$, we start from the ordering graph $J$ prior to the split of module $T$, and then update this graph to a new graph $J_{\mathrm{new}}$ that reflects the module order after splitting module $T$.
The order of modules $B$ and $C$ themselves (after splitting module $T$) reduces to splitting module $T$ into module $B$ and $C$ within a three-module case $A\rightharpoonup T$, and so can be determined using Rule~\ref{Ar5}.
The ordering of modules in the module groups $A$, $D$ and $E$ are unchanged by splitting module $T$, but the ordering of modules $B$ and $C$ and the other module groups can be determined using the following Rule.



\begin{RULE}[Identifying module ordering after splitting a module]\label{Ar6}
Suppose the module ordering graph $J$ prior to splitting module $T$ into modules $B$ and $C$ has nodes $\mathcal{V}(J)$ and edges $\mathcal{E}(J)$. The new ordering graph $J_{\mathrm{new}}$ after splitting module $T$ has nodes $(\mathcal{V}(J) \cup \{B, C\}) \setminus \{T\}$. The initial set of edges for $J_{\mathrm{new}}$ consists of all edges in $J$, except for all edges to or from module $T$.
Besides adding an edge between $B$ and $C$ if $B\rightharpoonup C$ or $C\rightharpoonup B$, additional edges are then added to $J_{\mathrm{new}}$ as follows, corresponding to potential links to module groups $A$, $D$ and $E$:
\begin{enumerate}
    \item For modules in $A$ we have $A\rightharpoonup T$ and thus $\Psi_{A\cap T}\neq\emptyset$. For all modules $A^\prime$ in module group $A$:
    \begin{enumerate}
        \item When module $A^\prime$ is not a parent module of $T$ in the original ordering graph $J_{\mathrm{original}}$, there is no direct relationship between $A^\prime$ and any module from $\{B,C\}$ and we do not add arrows between them in the ordering graph $J_{\mathrm{new}}$.
        \item When module $A^\prime$ is a parent module of $T$ in the original ordering graph $J_{\mathrm{original}}$ and suppose that $B$ is the parent module in ordering graph $J_{T}$:
        \begin{enumerate}
            \item If $\Psi_{A^\prime\cap B}\neq\emptyset$, we say that module $A^\prime$ is a parent module of module $B$ and we add an arrow from $A^\prime$ to $B$ in the ordering graph $J_{\mathrm{new}}$.
            \item If $\Psi_{A^\prime\cap B}=\emptyset$ but $\Psi_{A^\prime\cap C}\neq\emptyset$, we say that module $A^\prime$ is a parent module of module $C$ and we add an arrow from $A^\prime$ to $C$ in the ordering graph $J_{\mathrm{new}}$.
        \end{enumerate}
        \item When module $A^\prime$ is a parent module of $T$ in the original ordering graph $J_{\mathrm{original}}$ and suppose that $B$ and $C$ are unordered in the ordering graph $J_{T}$, given an arbitrary module $T^\prime$ from $\{B,C\}$:
        \begin{enumerate}
            \item If $\Psi_{A^\prime\cap T^\prime}\neq\emptyset$, we say that module $A^\prime$ is a parent module of module $T^\prime$ and we add an arrow from $A^\prime$ to $T^\prime$ in the ordering graph $J_{\mathrm{new}}$.
            \item Otherwise, there is no direct relationship between $A^\prime$ and module $T^\prime$ and we do not add an arrow between them in the ordering graph $J_{\mathrm{new}}$.
        \end{enumerate}
    \end{enumerate}
    \item For modules in $D$ we have $T\rightharpoonup D$ and thus $\Psi_{D\cap T}\neq\emptyset$. For all modules $D^\prime$ in module group $D$:
    \begin{enumerate}
        \item When module $D^\prime$ is not a child module of $T$ in the original ordering graph $J_{\mathrm{original}}$, there is no direct relationship between $D^\prime$ and any module from $\{B,C\}$ and we do not add arrows between them in the ordering graph $J_{\mathrm{new}}$.
        \item When module $D^\prime$ is a child module of $T$ in the original ordering graph $J_{\mathrm{original}}$ and suppose that $B$ is the child module in ordering graph $J_{T}$:
        \begin{enumerate}
            \item If $\Psi_{D^\prime\cap B}\neq\emptyset$, we say that module $D^\prime$ is a child module of module $B$ and we add an arrow from $B$ to $D^\prime$ in the ordering graph $J_{\mathrm{new}}$.
            \item If $\Psi_{D^\prime\cap B}=\emptyset$ but $\Psi_{D^\prime\cap C}\neq\emptyset$, we say that module $D^\prime$ is a child module of module $C$ and we add an arrow from $C$ to $D^\prime$ in the ordering graph $J_{\mathrm{new}}$.
        \end{enumerate}
        \item When module $D^\prime$ is a child module of $T$ in the original ordering graph $J_{\mathrm{original}}$ and suppose that $B$ and $C$ are unordered in the ordering graph $J_{T}$, given an arbitrary module $T^\prime$ from $\{B,C\}$:
        \begin{enumerate}
            \item If $\Psi_{D^\prime\cap T^\prime}\neq\emptyset$, we say that module $D^\prime$ is a child module of module $T^\prime$ and we add an arrow from $T^\prime$ to $D^\prime$ in the ordering graph $J_{\mathrm{new}}$.
            \item Otherwise, there is no direct relationship between $D^\prime$ and module $T^\prime$ and we do not add an arrow between them in the ordering graph $J_{\mathrm{new}}$.
        \end{enumerate}

    \end{enumerate}
     \item There is no direct relationship between any module $E^\prime$ in module group $E$ and modules $B$ or $C$, since the ordering of $E$ and $T$ is $(E,T)$ (i.e., $\Psi_{E\cap T}=\emptyset$). We thus do not add any edges to $J_{\mathrm{new}}$ either to or from modules in module group $E$.
\end{enumerate}
\end{RULE}


Given the new module order graph $J_{\mathrm{new}}$, the grouped module order graph can be obtained by Definition~\ref{DEgroupmoduleordergraph}, and depicted in Figure~\ref{F7}.

\subsubsection{Cut distribution}

The cut distribution after splitting module $T$ into modules $B$ and $C$ is given in grouped form by the following definition.

\begin{definition}[Cut distribution: sequential splitting case]
\label{DEsequentialcut}
Given a DAG $G=(\Psi,\mathcal{E})$ with modules $\Psi_A = (X_A,\Theta_A)$ and $\Psi_B = (X_B,\Theta_B)$ formed according to Rule~\ref{Ar1}, when the module order according to Rule~\ref{Ar2} is $A\rightharpoonup B$, the cut distribution for parameters $\Theta$ is:
\begin{enumerate}
    \item When module $B$ and $C$ are unordered conditional on $A$ and module $D$ is a child module of both $B$ and $C$ (Figure \ref{F7} (b)),
\begin{multline*}
p(\Theta\given X) = p(\Theta_{(A\cup B\cup C\cup D\cup E)^\mathsf{c}}\given \text{pa}(\Theta_{(A\cup B\cup C\cup D\cup E)^\mathsf{c}})) \\
\times p(\Theta_{D\setminus (A\cup B\cup C\cup E)}\given \Theta_{(A\cup B\cup C\cup E)\cap D}, X_D)\,p(\Theta_{E\setminus A}\given \Theta_{A\cap E}, X_E)\\
\times p(\Theta_{C\setminus A}\given \Theta_{A\cap C}, X_C)\,p(\Theta_{B\setminus A}\given \Theta_{A\cap B}, X_B)\,p(\Theta_A\given X_A).
\end{multline*}

\item When module $B$ and $C\cup D$ are unordered conditional on $A$ (Figure \ref{F7} (c)),
\begin{multline*}
p(\Theta\given X) = p(\Theta_{(A\cup B\cup C\cup D\cup E)^\mathsf{c}}\given \text{pa}(\Theta_{(A\cup B\cup C\cup D\cup E)^\mathsf{c}})) \\
\times p(\Theta_{D\setminus (A\cup C\cup E)}\given \Theta_{(A\cup C\cup E)\cap D}, X_D)  p(\Theta_{E\setminus A}\given \Theta_{A\cap E}, X_E)\\
\times p(\Theta_{C\setminus A}\given \Theta_{A\cap C}, X_C)\,p(\Theta_{B\setminus A}\given \Theta_{A\cap B}, X_B)\,p(\Theta_A\given X_A),
\end{multline*}
where the component of module $D$:
\[
p(\Theta_{D\setminus (A\cup C\cup E)}\given \Theta_{(A\cup C\cup E)\cap D}, X_D)
\]
is still unchanged because $\Theta_{B\cap D}=\emptyset$ (this is also the key difference between this case and previous case).

\item When module $A\cup B$ and $C$ are unordered (Figure \ref{F7} (d)),
\begin{multline*}
p(\Theta\given X) = p(\Theta_{(A\cup B\cup C\cup D\cup E)^\mathsf{c}}\given \text{pa}(\Theta_{(A\cup B\cup C\cup D\cup E)^\mathsf{c}}))\\
\times p(\Theta_{D\setminus (A\cup B\cup C\cup E)}\given \Theta_{(A\cup B\cup C\cup E)\cap D}, X_D)\,p(\Theta_{E\setminus A}\given \Theta_{A\cap E}, X_E) \\
\times p(\Theta_C\given X_C)\,p(\Theta_{B\setminus A}\given \Theta_{A\cap B}, X_B)\, p(\Theta_A\given X_A).
\end{multline*}

\item When module $A\cup B$ and $C\cup D$ are unordered (Figure \ref{F7} (e)),
\begin{multline*}
p(\Theta\given X) = p(\Theta_{(A\cup B\cup C\cup D\cup E)^\mathsf{c}}\given \text{pa}(\Theta_{(A\cup B\cup C\cup D\cup E)^\mathsf{c}})) \\
\times p(\Theta_{D\setminus (A\cup C\cup E)}\given \Theta_{(A\cup C\cup E)\cap D}, X_D)\,p(\Theta_{E\setminus A}\given \Theta_{A\cap E}, X_E)\\
\times p(\Theta_C\given X_C)\,p(\Theta_{B\setminus A}\given \Theta_{A\cap B}, X_B) \, p(\Theta_A\given X_A).
\end{multline*}

\item When module $A\cup B\cup D$ and $C$ are unordered (Figure \ref{F7} (f)),
\begin{multline*}
p(\Theta\given X) = p(\Theta_{(A\cup B\cup C\cup D\cup E)^\mathsf{c}}\given \text{pa}(\Theta_{(A\cup B\cup C\cup D\cup E)^\mathsf{c}})) \\
\times p(\Theta_{D\setminus (A\cup B\cup E)}\given \Theta_{(A\cup B\cup E)\cap D}, X_D)\,p(\Theta_{E\setminus A}\given \Theta_{A\cap E}, X_E)\\
\times p(\Theta_C\given X_C)\,p(\Theta_{B\setminus A}\given \Theta_{A\cap B}, X_B)\,p(\Theta_A\given X_A).
\end{multline*}

\item When module $A$ and $B$ are unordered (Figure \ref{F7} (g)),
\begin{multline*}
p(\Theta\given X) = p(\Theta_{(A\cup B\cup C\cup D\cup E)^\mathsf{c}}\given \text{pa}(\Theta_{(A\cup B\cup C\cup D\cup E)^\mathsf{c}}))  \\
\times p(\Theta_{D\setminus (A\cup B\cup C\cup E)}\given \Theta_{(A\cup B\cup C\cup E)\cap D}, X_D)  p(\Theta_{E\setminus A}\given \Theta_{A\cap E}, X_E)\\
\times  p(\Theta_{C\setminus (A\cup B)}\given \Theta_{(A\cup B)\cap C}, X_C)\,p(\Theta_B\given  X_B)\,p(\Theta_A\given X_A).
\end{multline*}

\item When module $C$ is the child module of $B$ (Figure \ref{F7} (h)),
\begin{multline*}
p(\Theta\given X) = p(\Theta_{(A\cup B\cup C\cup D\cup E)^\mathsf{c}}\given \text{pa}(\Theta_{(A\cup B\cup C\cup D\cup E)^\mathsf{c}})) \\
\times p(\Theta_{D\setminus (A\cup B\cup C\cup E)}\given \Theta_{(A\cup B\cup C\cup E)\cap D}, X_D)  p(\Theta_{E\setminus A}\given \Theta_{A\cap E}, X_E) \\
\times p(\Theta_{C\setminus (A\cup B)}\given \Theta_{(A\cup B)\cap C}, X_C)
 p(\Theta_{B\setminus A}\given \Theta_{A\cap B}, X_B)\,p(\Theta_A\given X_A).
\end{multline*}

\item When module $C$ and $D$ are unordered conditional on module $B$ (Figure \ref{F7} (i)),
\begin{multline*}
p(\Theta\given X) = p(\Theta_{(A\cup B\cup C\cup D\cup E)^\mathsf{c}}\given \text{pa}(\Theta_{(A\cup B\cup C\cup D\cup E)^\mathsf{c}})) \\
\times p(\Theta_{D\setminus (A\cup B\cup E)}\given \Theta_{(A\cup B\cup E)\cap D}, X_D)\,p(\Theta_{E\setminus A}\given \Theta_{A\cap E}, X_E)\\
\times p(\Theta_{C\setminus (A\cup B)}\given \Theta_{(A\cup B)\cap C}, X_C)\,p(\Theta_{B\setminus A}\given \Theta_{A\cap B}, X_B)\,p(\Theta_A\given X_A).
\end{multline*}
\end{enumerate}
 \end{definition}

In ungrouped form, the cut distribution for the new ordering graph $J_{\mathrm{new}}$ is given by the following definition.

\begin{definition}
\label{DEcutorderinggraph}
The cut distribution corresponding to ordering graph $J$ for modules $\mathcal{M}$, with $\mathrm{an}_{J}(M)$ denoting the ancestor modules of module $M \in \mathcal{M}$.
\begin{equation*}
p_{J}(\Theta \given X)
=
\prod_{M \in \mathcal{M}}
p(\Theta_{M \setminus \mathrm{an}_{J}(M)} \given X_{M}, \Theta_{M \cap \mathrm{an}_{J}(M)})
\end{equation*}
\end{definition}

In summary, having identified the ordering of modules, the inference of any module can be only conducted after the inferences of all its ancestor modules are conducted.

\subsubsection{Sequential formation of cut distributions}

A general algorithm for modularising inference for a model with DAG $G=(\Psi,\mathcal{E})$ where $\Psi = (X, \Theta)$ can now be described.
We first partition the observable nodes $X$ into an arbitrary number of sets $
\{X^{\ast}_{A_{1}}, \dots, X^{\ast}_{A_{s}}\}$, ordered such that
$X^{\ast}_{A_{i}}$ is regarded as more reliable than $X^{\ast}_{A_{j}}$ for $i < j$. Thus, $X^{\ast}_{A_{1}}$ contains the observable variables regarded as most reliable, and $X^{\ast}_{A_{s}}$ the least reliable.
The cut distribution can then be formed using Algorithm~\ref{alg1}.

\begin{algorithm}[t]
\caption{Sequential formation of cut distributions}
\label{alg1}
\begin{algorithmic}[1]
\Require A partition of observable nodes $X$ into sets $
\{X^{\ast}_{A_{1}}, \dots, X^{\ast}_{A_{s}}\}$ ordered such that $X^{\ast}_{A_{i}}$ is regarded as more reliable than $X^{\ast}_{A_{j}}$ for $i < j$.

\State Apply Rule~\ref{Ar1} to $X^{\ast}_{A_{1}}$ and $X^{\ast}_{A_{2}}, \dots, X^{\ast}_{A_{s}}$ to identify the corresponding modules $M_{1}$ and $M_{(2, \dots, s)}$.
\State Apply Rule~\ref{Ar2} to determine the order of modules $M_{1}$ and $M_{(2, \dots, s)}$.

\State Apply Rule~\ref{Ar1} to $X^{\ast}_{A_{2}}$ and $X^{\ast}_{A_{3}}, \dots, X^{\ast}_{A_{s}}$ to identify the corresponding modules $M_2$ and $M_{(3, \dots, s)}$.
\State Apply Rule~\ref{Ar5} to determine the order of modules $M_1$, $M_2$ and $M_{(3, \dots, s)}$.
\State Set $J$ to the module order graph for $M_1$, $M_2$ and $M_{(3, \dots, s)}$

\For{$r = 3, \dots, s-1$}


\State Form module groups: $T = M_{(r, \dots, s)}$; ancestors $A$ and descendants $D$ of $T$; and other modules $E$.

\State Apply Rule~\ref{Ar1} to $X^{\ast}_{A_{r}}$ and $X^{\ast}_{A_{r+1}}, \dots, X^{\ast}_{A_{s}}$ to identify the corresponding modules $M_r$ and $M_{(r+1, \dots, s)}$.

\State Apply Rule~\ref{Ar5} to determine the order of the modules $M_{r}$, $M_{(r+1, \dots, s)}$ and $A$.

\State Apply Rule~\ref{Ar6} to form the new ordering graph $J_{\mathrm{new}}$, and set $J = J_{\mathrm{new}}$.
\EndFor
\State Apply Definition~\ref{DEcutorderinggraph} to form cut distribution $p_{J}(\Theta \given X)$
\end{algorithmic}
\end{algorithm}

\section{Illustrative examples}
In this section, we illustrate cut inference by applying Rules~\ref{Ar1}-\ref{Ar6} that we have proposed. We first discuss a particular two-module case in which a within-module cut is applied. (The standard, generic two-module case that has been considered by many previous studies is discussed in Appendix B of the Supplementary Materials.) We then discuss a particular longitudinal model in which there is misspecification and apply a multiple-module cut. We show that the derived cut distribution can reduce estimation bias compared to the standard posterior distribution.

\subsection{Two-module salmonella source attribution model}
We now consider a simplified version of the two-module salmonella source attribution model that has been studied in \textcite{https://doi.org/10.1111/risa.13310}. The model is simplified here to avoid complications that are unnecessary for illustration purposes. The purpose of the salmonella source attribution model is to quantify the proportion of human salmonella infection cases that are attributed to specific food sources in which specific salmonella subtypes exist.

Suppose there are $i=1,\dots,I$ predefined food sources and $s=1,\dots,S$ subtypes of salmonella, and let $t=1,\dots,T$ denote the time period. We assume the observed number of human salmonella infection cases $C_{s,t}$ due to subtype $s$ at time $t$ follows a Poisson distribution:
\begin{equation}
C_{s,t}\sim \textrm{Poisson}\left(\sum_{i=1}^I L_{i,t} \, r_{i,s,t}\, a_i\, q_s \right),
\label{E14}
\end{equation}
for $s=1,\dots,S$ and $t=1,\dots,T$, where $L_{i,t}$ is gross exposure of salmonella in food source $i$ at time $t$; $r_{i,s,t}$ is the relative proportion of salmonella subtype $s$ in food source $i$ at time $t$; the source-specific parameter $a_i$ accounts for differences between food sources in their ability to cause a human salmonella infection; and, similarly, the subtype-specific parameters $q_s$ account for differences in the subtypes.

The relative proportion of subtype $r_{i,s,t}$ is informed by a separate dataset, which records the observed annual counts $X_{i,s,t}$ of the subtypes $s$ in food source $i$ at time $t$. This is modelled as a multinomial distribution with respect to $r_{i,s,t}$ as:
\[
(X_{i,1,t},\dots, X_{i,S,t}) \sim \textrm{Multinomial}\left((r_{i,1,t},\dots, r_{i,S,t}), N_{i,t} \right),
\]
for $i=1,\dots,I$ and $t=1,\dots,T$, where $N_{i,t}=\sum_{s=1}^S r_{i,s,t}$ is the known total number of counts.

\begin{figure}[t]
\setlength{\abovecaptionskip}{0cm}
\setlength{\belowcaptionskip}{0cm}
\centering
\includegraphics[width=0.4\textwidth]{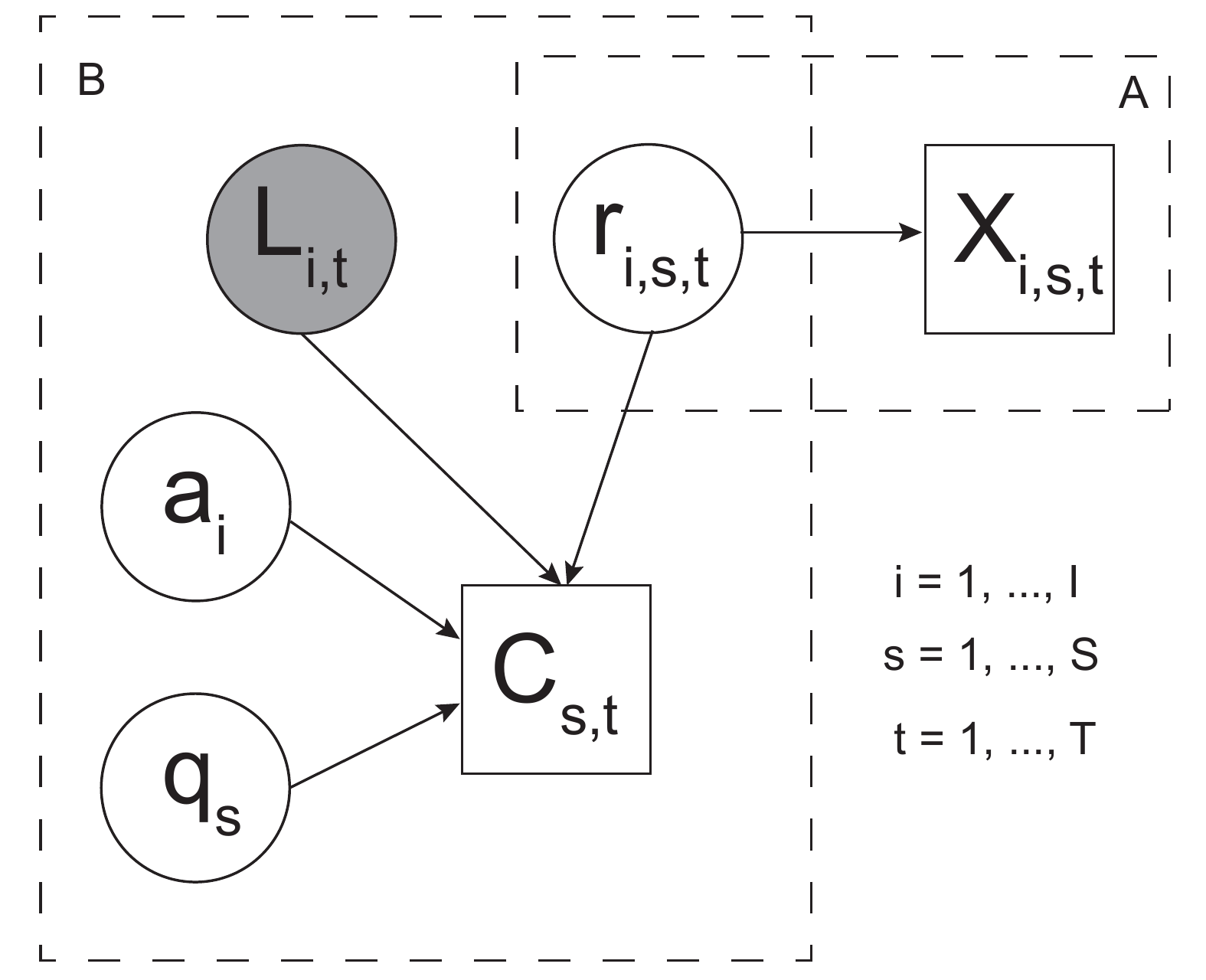}
\caption[DAG representation of the salmonella source attribution model.]{\textbf{DAG representation of the salmonella source attribution model.} The two modules are encapsulated by the dashed square. The parameter $L_{i,t}$ for which within-module cut inference is applied is coloured grey.}
\label{F9}
\end{figure}

A DAG representation for the model is shown in Figure \ref{F9}. We assume a Dirichlet prior for $(r_{i,1,t},\dots, r_{i,S,t})$; a Gamma prior for $L_{i,t}$; and exponential priors for both $a_i$ and $q_s$ for all $i$, $s$ and $t$. Given these priors, standard Bayesian inference can be applied via Bayes' theorem. However, \textcite{https://doi.org/10.1111/risa.13310} notes how this approach suffers from issues of identifiability between parameters $L_{i,t}$, $r_{i,s,t}$, $a_i$ and $q_s$ because the product of these parameters in \eqref{E14} (i.e., the expected number of human salmonella infection cases) can be the same for different combination of values of the four parameters.

This problem can be eased by applying cut inference so that part of the parameters are inferred separately, as described by \textcite{https://doi.org/10.1111/risa.13310}. After partitioning the observable random variables into two disjoint groups $X_A^\ast=\bigcup_{i,s,t =1 }^{I,S,T} X_{i,s,t}$ and $X_B^\ast= \bigcup_{s,t =1}^{S,T} C_{s,t}$, by Rule \ref{Ar1}, module $A$ is $\Psi_A=\bigcup_{i,s,t =1 }^{I,S,T}(r_{i,s,t},X_{i,s,t})$ and module $B$ is $\Psi_B=\bigcup_{i,s,t =1}^{I,S,T}(L_{i,t},a_i,q_s,r_{i,s,t},C_{s,t})$. Since $\Psi_{A\cap B}=\bigcup_{i,s,t =1 }^{I,S,T} r_{i,s,t}$, there are two directed edges: one from $\Psi_{A\cap B}$ to $\Psi_{A\setminus B}$ (i.e., $r_{i,s,t}\rightarrow X_{i,s,t}$); the other from $\Psi_{A\cap B}$ to $\Psi_{B\setminus A}$ (i.e., $r_{i,s,t}\rightarrow C_{s,t}$). Hence, by Rule \ref{Ar2}, we can choose either module $A$ or $B$ to be the parent module. Here, we choose to let module $A$ be the parent module so that the parameters $r_{i,s,t}$ can be inferred solely by information from module $A$ resulting in the following distribution for the parameters $r_{i,s,t}$:
\[
p(r_{i,s,t}\given X_{i,s,t}) \propto p(X_{i,s,t}\given r_{i,s,t})\,p(r_{i,s,t}).
\]
To further ease the identifiability issues, we follow \textcite{https://doi.org/10.1111/risa.13310} and apply a within-module cut on the parameter $L_{i,t}$. That is, we ensure that $L_{i,t}$ is inferred solely using its prior distribution $p(L_{i,t})$. Now the cut distribution is:
\begin{equation*}
p_{A\rightharpoonup B} \left(\bigcup_{i,s,t = 1}^{I,S,T} r_{i,s,t},L_{i,t}, a_i, q_s\right) = \prod_{i,s,t = 1}^{I,S,T} p(a_i, q_s\given r_{i,s,t},L_{i,t},C_{s,t})\, p(r_{i,s,t}\given X_{i,s,t})\,p(L_{i,t}).
\end{equation*}
Here, module $A$ is inferred solely via $p(r_{i,s,t}\given X_{i,s,t})$ and the within-module cut in module $B$ is invoked by inferring $L_{i,t}$ using only its prior distribution. All remaining parameters in module $B$ are inferred using conditional self-contained Bayesian inference:
\[
p(a_i, q_s\given r_{i,s,t},L_{i,t},C_{s,t}) \propto p(C_{s,t}\given r_{i,s,t},L_{i,t}, a_i, q_s)\,p(a_i)\,p(q_s).
\]
In summary, we note that the model constructed informally by \textcite{https://doi.org/10.1111/risa.13310} falls into the general framework that we have proposed.

\subsection{Misspecified longitudinal model} \label{SE3.3}
We now show that multiple-module cut inference can reduce estimation bias when there is misspecification in a particular Gaussian regression example involving longitudinal observations over a period of time. Models that involve longitudinal observations naturally fit the multiple-module case, since modules can be formed according to the timepoint. We will show the estimation bias of standard Bayesian inference due to misspecification and illustrate how this bias is reduced by adopting cut inference. A numerical simulation of this example is presented in Appendix C in the supplementary materials.

\begin{figure}[t]
\setlength{\abovecaptionskip}{0cm}
\setlength{\belowcaptionskip}{0cm}
\centering
\includegraphics[width=0.6\textwidth]{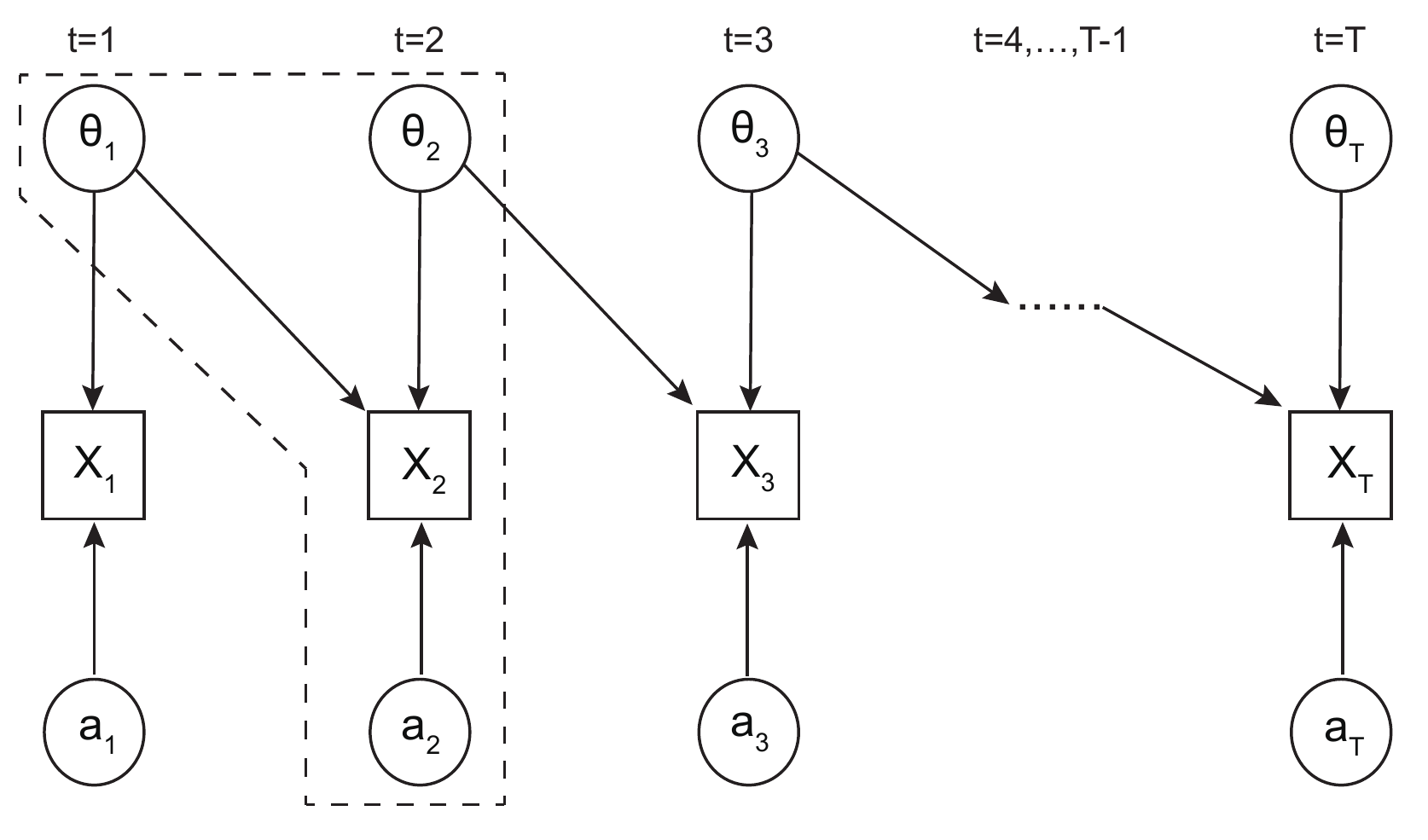}
\caption[DAG representation of the sequential Gaussian regression model.]{\textbf{DAG representation of the sequential Gaussian regression model.} The dashed area highlights a single self-contained Bayesian module $\Psi_2=(\theta_1,\theta_2,a_2,X_2)$.}
\label{F10}
\end{figure}

Suppose the true data generating process is according to the DAG as shown in Figure \ref{F10}. At each time $t=1,\dots,T$, we have an $n$-dimensional observation $X_t=(X_{t1},\dots,X_{tn})$ with the following true data generating process:
\begin{equation}
\begin{aligned}
X_1\given \nu_1 &\sim \mathcal{N}\left(P_1 \nu_1, \textbf{I} \right); \\
X_t\given \nu_t,\theta_{t-1} &\sim \mathcal{N}\left(P_t \nu_t+Q_t f^\ast(\theta_{t-1}), \textbf{I} \right), \qquad t = 2, 3, \dots
\end{aligned}
\label{E15}
\end{equation}
Here, $\theta_t$, $t=1,\dots,T$, is the parameter of interest and let $\nu_t = (a_t,\theta_t)^\mathsf{T}$, $t=1,\dots,T$ be the augmented parameter which includes an additional intercept term $a_t$; and $f^\ast(\cdot)$ is an unknown deterministic function in the true data generating process. The current covariate matrix $P_t$ is:
\[
P_t =
\setlength\arraycolsep{5pt}
\begin{pmatrix}
1 & p_{t1} \\
\vdots  & \vdots  \\
1 & p_{tn}
\end{pmatrix};
\]
and the past covariate vector is $Q_t=(q_{t1},\dots,q_{tn})$, $t = 2, 3, \dots$. We further assume that $p_{tk}$ and $q_{tk}$ are realizations of covariate variables, which have been drawn independently with mean 0. The 0 mean assumption is plausible because we can always centralize $P_t$ and $Q_t$ and this centralization can be compensated by the intercept parameter $a_t$.
The covariance matrix is assumed to be the identity matrix $\textbf{I}$ for simplicity.

We now suppose the true data generating process is partially unknown. Specifically, we assume the function $f^\ast(\cdot)$ is unknown, and that the analysis model is misspecified as $f(\cdot) = f^\ast(\cdot) + \delta$, where $\delta$ quantifies the degree of misspecification.

In this Section we will calculate and compare the estimation bias when using cut inference and using standard Bayesian inference. To make the cut distribution tractable, we use the conjugate prior $\nu_t\sim\mathcal{N}(0, \textbf{I})$.

\subsubsection{Cut inference}
We first consider applying cut inference and show that the estimation bias will approximately be 0 on an average basis with respect to realizations of $P_t$ and $Q_t$. The idea is that, at every time $t$, we cut the feedback from time $t+1$ so that misspecification beyond $t+1$ will not affect the inference before that time.

For now, assume that the misspecification of $f^\ast(\cdot)$ only exists between time $t-1$ and $t$ for a specific time $0<t<T$ (we will relax this assumption later).
To form modules, we first partition the observable random variables $X=(X_1,\dots,X_T) = X_{1:T}$ into two disjoint groups $X_A^\ast=X_{1:(t-1)}$ and $X_B^\ast=X_{t:T}$. Then by Rule \ref{Ar1}, we have two self-contained Bayesian modules $\Phi_A=\{\theta_{1:(t-1)},\allowbreak a_{1:(t-1)},\allowbreak X_{1:(t-1)}\}$ and $\Phi_B=\{\theta_{(t-1):T},a_{t:T},X_{t:T}\}$. Because the misspecification relates to the function $f^\ast(\cdot)$ which links observations at time $t-1$ and $t$, we wish to prevent information from times $t, t+1, \dots, T$ being used, which will avoid bias for inference before time $t$. Hence, although by Rule~\ref{Ar2} both orders hold, we choose module $B$ to be the child module.

First, note there is no misspecification before time $t-1$ and thus a standard Bayesian inference for $(\nu_1,\dots,\nu_{t-1})$ with information only from module $A$ can be conducted. This is available because module $A$ is a self-contained Bayesian module. Second, to calculate the estimation bias at time $t$, we split module $B$ into two modules $\Psi_C=\{\theta_{t-1},\theta_t,a_t,X_t\}$ and $\Phi_D=\{\theta_{t:T},a_{(t+1):T},X_{(t+1):T}\}$ and similarly treat module $D$ as the child module of module $C$ (i.e., we have the order $A\rightharpoonup C\rightharpoonup D$). To simplify the notation, we write
\[
\Lambda_t = P_t^\mathsf{T} P_t + \textbf{I} =
\begin{pmatrix}
n+1 & \sum p_{ti} \\
\sum p_{ti} & \sum p_{ti}^2+1
\end{pmatrix}
\]
Now by Definition \ref{DE3} (case: $A\rightharpoonup C\rightharpoonup D$), the conditional self-contained Bayesian posterior of $\nu_t$, conditional on $\theta_{t-1}$, is
\[
\nu_t\given X_t,\theta_{t-1} \sim \mathcal{N}\left( \Lambda_t^{-1} P_t^\mathsf{T} (X_t-Q_tf(\theta_{t-1})),\ \Lambda_t^{-1} \right),
\]
where $f(\cdot)$ is the misspecified function such that $f(\cdot) = f^\ast(\cdot) + \delta$, and thus
\begin{equation}
\nu_t\given X_t,\theta_{t-1} \sim \mathcal{N}\left(\Lambda_t^{-1} P_t^\mathsf{T} (X_t-Q_tf^\ast(\theta_{t-1})) - \Lambda_t^{-1} P_t^\mathsf{T}Q_t\delta,\ \Lambda_t^{-1}\right).
\label{E16}
\end{equation}
It is clear that the estimation bias is $\Delta_t=K_t \delta$, where $K_t:=\Lambda_t^{-1} P_t^\mathsf{T}Q_t$. Since
\[
\Lambda_t^{-1} = \frac{1}{(n+1)(\sum p_{ti}^2+1)-(\sum p_{ti})^2}
\begin{pmatrix}
\sum p_{ti}^2 +1 & -\sum p_{ti} \\
-\sum p_{ti} & n+1
\end{pmatrix},
\]
the estimation bias for the intercept $a=(a_1,\dots,a_T)$ is
\[
K_{t1} = \frac{(\sum p_{ti}^2 +1) \sum q_{ti} - \sum p_{ti} \sum p_{ti}q_{ti}}{(n+1)(\sum p_{ti}^2+1)-(\sum p_{ti})^2};
\]
and the estimation bias for the parameter of interest $(\theta_1,\dots,\theta_T)$ is
\[
K_{t2} = \frac{(n+1) \sum p_{ti}q_{ti} - \sum p_{ti} \sum q_{ti}}{(n+1)(\sum p_{ti}^2+1)-(\sum p_{ti})^2}.
\]
Both $K_{t1}$ and $K_{t2}$ have mean 0 so the estimation bias will distribute around 0.

Now we consider the case when the function $f^\ast(\cdot)$ is misspecified for all $t$ meaning that we would like to cut feedback between all times $t$ and $t+1$. We therefore recursively split the modules until each module involves a single observable random variable. In other words we choose the partition $\{X_{A_1}^\ast, \dots, {X_{A_T}}^\ast\}$ of the observable random variables with $X_{A_t}^\ast = X_t$ for $t=1,\dots,T$.
Applying Algorithm~\ref{alg1}, module $M_1$ is $\Psi_1=\{\theta_1,a_1,X_1\}$ and module $M_{(2, \dots, T)} = \{\theta_{1:T},a_{2:T},X_{2:T}\}$, and while both orderings hold by Rule~\ref{Ar2}, we choose $M_1 \rightharpoonup M_{(2, \dots, T)}$ for the same reason as in the two-module case.
To split $M_{(2, \dots, T)}$, we first use Rule~\ref{Ar1} that identifies module $M_2 =\{\theta_{1:2},a_2,X_2\}$ and $M_{(3, \dots, T)} = \{\theta_{2:T},a_{3:T},X_{3:T}\}$.
Since we are splitting the child module $M_{(2, \dots, T)}$, and $\Psi_{M_{1} \cap M_{2}} = \{\theta_{1}\}\neq \emptyset$ but $\Psi_{M_{1} \cap M_{(3, \dots, T)}} = \emptyset$, Rule~\ref{Ar5}.1(b) applies.
There are edges from $\Psi_{\underline{M_{2}M_{(3, \dots, T)}}} = \{\theta_2\}$ to both $\Psi_{\underline{M_{2}}} \cup X_2^\ast = \{a_2, X_2\}$ and $\Psi_{\underline{M_{(3, \dots, T)}}} \cup X_{(3, \dots, T)}^\ast = \{\theta_{3:T}, a_{3:T}, X_{3:T}\}$, so both $M_{1} \rightharpoonup M_{2} \rightharpoonup M_{(3, \dots, T)}$ and $(M_{1}, M_{(3, \dots, T)}) \rightharpoonup M_{2}$ hold.
We choose $M_{1} \rightharpoonup M_{2} \rightharpoonup M_{(3, \dots, T)}$.
To split $M_{(3, \dots, T)}$ into $M_{3} = \{\theta_{2:3}, a_3, X_3\}$ and $M_{4} = \{\theta_{3:T}, a_{4:T}, X_{4:T}\}$, we consider its ancestor module group $A = \{M_{1}, M_{2}\}$; there has no descendants or other modules.
Similarly to the previous split, by Rule~\ref{Ar5} both $A \rightharpoonup M_{3} \rightharpoonup M_{(4, \dots, T)}$ and $(A, M_{3}) \rightharpoonup M_{(4, \dots, T)}$ hold, but we choose the former, and using Rule~\ref{Ar6}.1(a) and Rule~\ref{Ar6}.1(b)(i) we find the new ordering is $M_{1} \rightharpoonup M_{2} \rightharpoonup M_{3} \rightharpoonup M_{(4, \dots, T)}$. Recursively applying the same argument we ultimately find that $M_{1} \rightharpoonup M_{2} \rightharpoonup \dots \rightharpoonup M_{T}$, with $M_{t} = \{\theta_{(t-1):t}, a_t, X_t\}$ for $t=2, \dots T$.
Note that at each time $t$ estimation bias could still accumulate for future modules (i.e., descendant modules). Denote $\theta_t^\ast$ as the true value of $\theta_t$ and $\nu_t^\ast$ as the true value of $\nu_t$ for any $t$. Consider the estimation bias at time $t=2$, based on \eqref{E16}. Since module $\Psi_1$ does not involve any misspecification, the conditional self-contained Bayesian posterior of $\nu_1$ has mean $\nu_1^\ast$, we can write
\[
\mathbb{E}_{\theta_1}\left(\mathbb{E}(\nu_2 \given  X_2, \theta_1)\right)= \nu_2^\ast - K_2\delta.
\]
Then based on \eqref{E16}, it is easy to deduce that
\begin{align*}
\mathbb{E}(\nu_3 \given  X_3, \theta_2) &= \Lambda_3^{-1} P_3^\mathsf{T} (X_3-Q_3f^\ast(\theta_2)) - K_3\delta; \\
\mathbb{E}_{\theta_2} \left( \mathbb{E}(\nu_3 \given  X_3, \theta_2) \right) &= \mathbb{E}_{\theta_2} \left( \Lambda_3^{-1} P_3^\mathsf{T} (X_3-Q_3f^\ast(\theta_2)) \right) - K_3\delta \\
\ &\approx \mathbb{E}_{\theta_2} \left( \Lambda_3^{-1} P_3^\mathsf{T} \lbrace X_3-Q_3 [f^\ast(\theta_2^\ast)+ {f^\ast}^\prime(\theta_2^\ast)(\theta_2-\theta_2^\ast) ] \rbrace \right) - K_3\delta \\
\ &= \left( \Lambda_3^{-1} P_3^\mathsf{T} ( X_3-Q_3 f^\ast(\theta_2^\ast) ) \right) + \Lambda_3^{-1} P_3^\mathsf{T} Q_3 {f^\ast}^\prime(\theta_2^\ast) K_{22}\delta - K_3\delta \\
\ &= \left( \Lambda_3^{-1} P_3^\mathsf{T} ( X_3-Q_3 f^\ast(\theta_2^\ast) ) \right) - K_3 (1-{f^\ast}^\prime(\theta_2^\ast) K_{22})\delta.
\end{align*}
Hence, the estimation bias, up to module 3, is approximately equal to $K_3 (1-{f^\ast}^\prime(\theta_2^\ast) K_{22})\delta$. Similarly, we have
\begin{multline*}
\mathbb{E}_{\theta_2}\dots \mathbb{E}_{\theta_{t-1}} \left( \mathbb{E}(\nu_t \given  X_t, f(\theta_{t-1})) \right) \approx \left( \Lambda_t^{-1} P_t^\mathsf{T} ( X_t-Q_t f^\ast(\theta_{t-1}^\ast) ) \right) \\
- K_t \left(1- \sum_{i=1}^{t-2} \lbrace (-1)^{i+1} \prod_{j=t-i}^{t-1} [ {f^\ast}^\prime(\theta_j^\ast) K_{j2} ] \rbrace \right)\delta
\end{multline*}
Comparing with the original estimation bias $\Delta_t = K_t \delta$ when accumulation of bias is not considered, we observe that the estimation bias is perturbed by a multiplier $(1-\Upsilon_{t-1})$, where
\[
\Upsilon_{t-1} := \sum_{i=1}^{t-2} \lbrace (-1)^{i+1} \prod_{j=t-i}^{t-1} [ {f^\ast}^\prime(\theta_j^\ast) K_{j2} ] \rbrace.
\]
Note that $(1-\Upsilon_{t-1})$ has mean 1 since $K_{j2}$ has mean 0 for all $1\leq j\leq T$ and they are independent with each other. Hence, the mean (in terms of $P_t$ and $Q_t$) of the estimation bias is still 0 when function $f^\ast(\cdot)$ is misspecified for all $t$.

\subsubsection{Standard Bayesian inference with estimation bias}
Now we derive the standard posterior distribution for this model. Suppose for now that we perform inference under the correctly specified model. We can approximate the true data generating process \eqref{E15}, for any $t>1$, by Taylor approximation as a function of $\theta_{t-1}^\star$, which is a known constant value that is close to the true value $\theta_{t-1}$:
\[
X_t\given \nu_t,\theta_{t-1} \approxsim \mathcal{N}\left(P_t \nu_t+Q_t f^\ast(\theta_{t-1}^\star)+Q_t {f^\ast}^\prime(\theta_{t-1}^\star)(\theta_{t-1}-\theta_{t-1}^\star), \textbf{I} \right).
\]
Writing $\tilde{X}_t:= X_t - Q_t f^\ast(\theta_{t-1}^\star)+Q_t {f^\ast}^\prime(\theta_{t-1}^\star)\theta_{t-1}^\star$ and $m_{t-1}^\ast={f^\ast}^\prime(\theta_{t-1}^\star)$ for $t>1$, this can be simplified as
\[
\tilde{X}_t\given \nu_t,\theta_{t-1} \approxsim \mathcal{N}\left( P_t \nu_t+Q_t m_{t-1}^\ast \theta_{t-1}, \textbf{I} \right).
\]
Let $0_{n\times 2}$ denote a 0 matrix with $n$ rows and $2$ columns and let
\[
\tilde{Q}_t =
\begin{pmatrix}
0 & p_{t1} \\
\vdots & \vdots \\
0 & p_{tn}
\end{pmatrix}
\]
Then we can write the mean of $\tilde{X}$ in the matrix form as:
\[
\begin{aligned}
&\mathbb{E}\left[\begin{pmatrix}
\tilde{X}_1 \\
\tilde{X}_2 \\
\tilde{X}_3 \\
\vdots \\
\tilde{X}_T
\end{pmatrix}\right] =
\mathbb{E}(\tilde{X})\approx B \nu
=
\begin{pmatrix}
P_1 & 0_{n\times 2} & 0_{n\times 2} & \cdots & 0_{n\times 2} & 0_{n\times 2} \\
\tilde{Q}_2 m_1^\ast & P_2 & 0_{n\times 2} & \cdots & 0_{n\times 2} & 0_{n\times 2} \\
0_{n\times 2} & \tilde{Q}_3 m_2^\ast & P_3 & \cdots & 0_{n\times 2} & 0_{n\times 2} \\
\vdots & \vdots & \vdots & \ddots &\vdots & \vdots \\
0_{n\times 2} & 0_{n\times 2} & 0_{n\times 2} & \cdots & \tilde{Q}_T m_{T-1}^\ast & P_T
\end{pmatrix}
\begin{pmatrix}
\nu_1 \\
\nu_2 \\
\nu_3 \\
\vdots \\
\nu_T
\end{pmatrix}
\end{aligned}
\]
The standard posterior distribution of $\nu$ under the conjugate prior, when there is no misspecification (i.e., when $f^\ast(\cdot)$ is used), is
\[
\nu\given \tilde{X} \approxsim \mathcal{N}(\Lambda^{-1} B^\mathsf{T} \tilde{X}, \Lambda),
\]
where $\Lambda=(B^\mathsf{T} B + I)$ and can be interpreted as a block matrix
\[
\begin{pmatrix}
\Lambda_{1}^{(1)} & \Lambda_{2}^{(2)} & 0_{2\times 2} & \cdots & 0_{2\times 2} \\
\Lambda_{2}^{(2)} & \ddots & \ddots & \ddots & \vdots \\
0_{2\times 2} & \ddots & \ddots & \ddots & 0_{2\times 2} \\
\vdots & \ddots & \ddots & \ddots & \Lambda_{T}^{(2)} \\
0_{2\times 2} & \cdots & 0_{2\times 2} & \Lambda_{T}^{(2)} & \Lambda_{T}^{(1)}
\end{pmatrix}
\]
where
\begin{align*}
\Lambda_{t}^{(1)} &= \begin{pmatrix}
n+1 & \sum_{i=1}^n p_{ti} \\
\sum_{i=1}^n p_{ti} & \sum_{i=1}^n ( p_{ti}^2 + (m_t^\ast q_{(t+1)i} )^2) + 1
\end{pmatrix} \ \ \ \text{for}\ t<T\\
\Lambda_{T}^{(1)} &= \begin{pmatrix}
n+1 & \sum_{i=1}^n p_{Ti} \\
\sum_{i=1}^n p_{Ti} & \sum_{i=1}^n p_{Ti}^2 + 1
\end{pmatrix}\\
\Lambda_{t}^{(2)} &=
\begin{pmatrix}
  0 & 0 \\
  0 & m_{t-1}^\ast \sum_{i=1}^n (p_{ti}q_{ti})
\end{pmatrix}
\quad \text{for } t > 1;
\end{align*}
However, the model that we actually use is misspecified, with $f(\cdot) = f^\ast(\cdot) + \delta$. Under this misspecification, the standard posterior distribution is:
\[
\nu\given \tilde{X} \approxsim \mathcal{N} \left(\Lambda^{-1} B^\mathsf{T} \tilde{X} - \Lambda^{-1} B^\mathsf{T}
\begin{pmatrix}
0_{n\times 1} \\
Q_2 \\
\vdots \\
Q_T
\end{pmatrix}
\delta, \Lambda \right),
\]
where
\[
B^\mathsf{T}
\begin{pmatrix}
0_{n\times 1} \\
Q_2 \\
\vdots \\
Q_T
\end{pmatrix} =
\begin{pmatrix}
0 \\
\sum (q_{2i}m_1^\ast)^2 \\
m_1^\ast \sum q_{2i} \\
m_1^\ast \sum p_{2i}q_{2i} + \sum (q_{3i}m_2^\ast)^2 \\
\vdots \\
m_{T-1}^\ast \sum q_{Ti} \\
m_{T-1}^\ast \sum p_{Ti}q_{Ti}.
\end{pmatrix}
\]
Unlike when using cut inference, there are no guarantees that the estimation bias $\Delta$ defined as
\[
\Delta :=
\Lambda^{-1} B^\mathsf{T}
\begin{pmatrix}
0_{n\times 1} \\
Q_2 \\
\vdots \\
Q_T
\end{pmatrix}
\delta
\]
has a mean (in terms of $P_t$ and $Q_t$) 0. Therefore, the estimation bias may not distribute around 0. An example of this is shown in the numerical simulation in Appendix C.

\section{Conclusion}
In this paper, we have formulated rules that describe how to apply cut inference within a modularised Bayesian framework. In particular, we formally define a ``module'', which we refer to as a ``self-contained Bayesian module'', as the set of variables, associated with a set of observable random variables, that can be used to conduct standard Bayesian inference for the module. We provide a rule prescribing how to form a self-contained Bayesian module from the structure of the DAG, starting from a partition of observable random variables. We describe the relationship between modules as a ``parent-child ordering'' in which inference of a child module is affected by its parent module (ancestor modules in multiple-module case) according to conditional self-contained Bayesian inference. This order can be identified directly from the DAG. Finally, we formulate the associated cut distributions, spanning from the basic two-module case to a more complicated multiple-module case. The extension of cut inference from the two-module case to the multiple-module case is available via our sequential splitting technique.

We define a self-contained Bayesian module in terms of the observable random variables rather than the parameters. There is a proposal of the definition of the cut inference in the two-module case via a variational message passing sampling scheme that concentrates on the parameters instead \parencite{yu2021variational}. The idea is that the factorized joint distribution \eqref{E8} can be viewed as a product of factors that involve parameters. For example, a likelihood $p(\theta\given \mathrm{pa}(\theta))$ can be regarded as a factor $f(\theta,\Theta\cap \mathrm{pa}(\theta))$ which involves parameter $\theta$ and $\Theta\cap \mathrm{pa}(\theta)$. The standard variational message passing algorithm for a particular $\theta$ from its full conditional density $p(\theta\given (\Theta\setminus \theta))$ which is proportional to the product of all factors that involve this $\theta$. To conduct cut inference for this $\theta$ when a particular factor is misspecified, they derive a new modified conditional density $p_{\text{cut}}(\theta\given (\Theta\setminus \theta))$ for $\theta$ by dropping the misspecified factor in the product. This proposal for cut inference is inspiring because it focuses on the parameter of interest and we envision it working well for the generic two-module case. However, a drawback of this approach is that the cut distribution is specified implicitly without reference to a DAG representation, making the effect of dropping factors more challenging to understand. This makes direct extension of both the proposed algorithm and the underlying definition to the multiple-module case unclear. It is also not clear how to handle the varying degrees of partial misspecification across modules (or in the factors according to the definition of \textcite{yu2021variational}) which is common in a multiple-module case. Furthermore, it is unclear how to determine in which order to drop factors when there are multiple misspecified factors. Nevertheless, we expect more discussions of this definition and studies of possible connections to our definition in the future.

Studies of cut inference within the generic two-module case have drawn broad attention recently. These include studies of the sampling methods for cut distribution \parencite{plummer2015cuts,https://doi.org/10.1111/rssb.12336,yu2021variational,liu2020stochastic}, the selection between cut distribution and standard posterior distribution \parencite{jacob2017better}, the asymptotic properties of cut distribution \parencite{pompe2021asymptotics}, the extension of cut inference for likelihood-free inference \parencite{chakraborty2022modularized}, the extension of cut inference for generalized Bayesian inference \parencite{https://doi.org/10.48550/arxiv.2202.09968} and the application of cut distributions in semi-parametric hidden Markov models \parencite{https://doi.org/10.48550/arxiv.2203.06081}. These studies are based on the generic two-module case, and extension of their results to the general multiple-module case remains unclear and may not be straightforward. We hope that this study can conceptualize cut inference for a broader range of statistical models, and enlighten future developments of methodology and algorithm, and stimulate applications of modularised Bayesian inference.

\section{Competing interests}
No competing interest is declared.

\section{Acknowledgments}
We thank the referees and associate editor whose feedback led to substantial improvements in presentation. We also thank Geoff Nicholls and Simon White for their feedback on an earlier version of this manuscript. Yang Liu was supported by a Cambridge International Scholarship from the Cambridge Commonwealth, European and International Trust. Robert J. B. Goudie was funded by the UK Medical Research Council [programme codes MC\_UU\_00002/2, MC\_UU\_00002/20 and MC\_UU\_00040/04] and The Alan Turing Institute.


\clearpage

\setcounter{section}{0}
\renewcommand{\thesection}{\Alph{section}}

\section{Supplementary Materials: Proofs}\label{appA}
\subsection{Proof of Lemma 1}
\begin{enumerate}
    \item We prove each point separately.

    \begin{itemize}
        \item Given Rule 1, it is straightforward to conclude that $\Psi_{(A\cup B)^\mathsf{c}}$ only contains parameters. This is because all observables will be in either $X_A$ or $X_B$ (or both) since $X = X_A^\ast \cup X_B^\ast$ is a partition of the observables.

    \item Now, suppose there is an edge $\psi_{(A\cup B)^\mathsf{c}} \rightarrow x_{A\cup B}$, where $x_{A\cup B}\in X_{A\cup B}$ and $\psi_{(A\cup B)^\mathsf{c}}\in \Psi_{(A\cup B)^\mathsf{c}}$. By Rule 1, we must incorporate $\psi_{(A\cup B)^\mathsf{c}}$ into either module $A$ or module $B$. This contradicts the fact that $\psi_{(A\cup B)^\mathsf{c}}\in \Psi_{(A\cup B)^\mathsf{c}}$.

    Now suppose there is an edge $\psi_{(A\cup B)^\mathsf{c}} \rightarrow \theta_{A\cup B}$, where $\theta_{A\cup B}\in \Theta_{A\cup B}$ and $\psi_{(A\cup B)^\mathsf{c}}\in \Psi_{(A\cup B)^\mathsf{c}}$. By Rule 1, $\theta_{A\cup B}$ must have a descendant $x^\ast\in X$ such that there is at least one directed path between $x^\ast=\psi_1,\cdots,\psi_s=\theta_{A\cup B}$ with $x^\ast$ as the leaf and every $\psi_l\in\Theta$, $l>1$. Then by Rule 1 we must incorporate $\psi_{(A\cup B)^\mathsf{c}}$ into the module associated with this $x^\ast$. This contradicts the fact that $\psi_{(A\cup B)^\mathsf{c}}\in \Psi_{(A\cup B)^\mathsf{c}}$.

    Since $\Psi_{A \cup B} = X_{A \cup B} \cup \Theta_{A \cup B}$ we conclude that $\text{ch}(\Psi_{(A\cup B)^\mathsf{c}})=\emptyset$.
    \end{itemize}

    \item Let $\psi_{A\setminus B}\leftarrow \psi^\ast$ be an edge where $\psi_{A\setminus B}\in\Psi_{A\setminus B}$ and $\psi^\ast\notin\Psi_{A\setminus B}$. By the \thmcrossref{first statement of Lemma 1}, $\psi^\ast\notin\Psi_{(A\cup B)^\mathsf{c}}$. By Rule 1, we must incorporate $\psi^\ast$ into module $A$, so it must belong to $\Psi_{A\cap B}$.

    Now let $\psi_{A\setminus B}\rightarrow \psi^\ast$ be an edge where $\psi_{A\setminus B}\in\Psi_{A\setminus B}$ and $\psi^\ast\notin\Psi_{A\setminus B}$. If $\psi^\ast\in\Psi_{B\setminus A}$, by Rule 1 we must incorporate $\psi_{A\setminus B}$ into module $B$. This has contradicted the fact that $\psi_{A\setminus B}\in\Psi_{A\setminus B}$. Hence, $\psi^\ast$ can only belong to either $\Psi_{A\cap B}$ or $\Psi_{(A\cup B)^\mathsf{c}}$.

    \item
    We consider the possibility that $\psi_{A \cup B}$ is a member of $X_{A \cap B}$ and $\Theta_{A \cap B}$ separately.

    Consider a V-structure $\psi_{A\setminus B}\rightarrow x_{A\cap B} \leftarrow \psi_{B\setminus A}$ where $x_{A\cap B}\in X_{A\cap B}$. Now $x_{A\cap B}$ must belong to either $X_A^\ast$ or $X_B^\ast$ since these are a partition of $X$. Suppose $x_{A\cap B} \in X_A^\ast$, then by Rule 1, we must incorporate both $\psi_{A\setminus B}$ and $\psi_{B\setminus A}$ into module A, which leads to a contradiction. The corresponding argument applies if $x_{A\cap B} \in X_B^\ast$.

    If there is a V-structure $\psi_{A\setminus B}\rightarrow \theta_{A\cap B} \leftarrow \psi_{B\setminus A}$ where $\theta_{A\cap B}\in \Theta_{A\cap B}$, then $\theta_{A\cap B}$ must have a descendant $x_A\in X_A^\ast$ such that there is at least one directed path between $\theta_{A\cap B}=\psi_1,\cdots,\psi_s=x_A$ with $x_A$ as the leaf and every $\psi_l\in\Theta$, $l<s$. By Rule 1 we must incorporate $\psi_{A\setminus B}$ and $\psi_{B\setminus A}$ into module $A$ and this leads to a contradiction.

    \item We first consider $\text{pa}(\Theta_{A\cap B})$. By Rule 1, any parent node $\psi\in \text{pa}(\Theta_{A\cap B})$ must also have descendants both in $X_A^\ast$ and $X_B^\ast$. Therefore, this $\psi$ belongs to both module $A$ and $B$. Because $\psi\notin \Theta_{A\cap B}$ by the definition of $\text{pa}(\cdot)$, then $\psi\in X_{A\cap B}$.

    We then consider $\text{pa}(X_{A\cap B})$. By the \thmcrossref{first statement of Lemma 1}, $\text{pa}(X_{A\cap B})\cap \Psi_{(A\cup B)^\mathsf{c}} = \emptyset$. Then $\text{pa}(X_{A\cap B})\subseteq \Psi\setminus\left(\Psi_{(A\cup B)^\mathsf{c}}\cup X_{A\cap B}\right)$. It is clear that $\Psi\setminus\left(\Psi_{(A\cup B)^\mathsf{c}}\cup X_{A\cap B}\right) = \{\Theta_{A\cap B}, \Theta_{A\setminus B}, \Theta_{B\setminus A}, X_{A\setminus B},X_{B\setminus A}\}$.

    \item We first consider the case when $\Psi_{(A\cup B)^\mathsf{c}}$-cut-sub-graph $G_{\text{cut}(\Psi_{(A\cup B)^\mathsf{c}})}$ has two disconnected components which are formed separately by nodes $\Psi_A$ and $\Psi_B$. If $\Psi_{A\cap B}\neq \emptyset$, then there must exist at least one $\psi\in\Psi_{A\cap B}$ which is the common ancestor of $X_A^\ast\subseteq \Psi_A$ and $X_B^\ast\subseteq \Psi_B$ by Rule 1. Hence $\Psi_A$ is connected with $\Psi_B$ which leads to a contradiction.

    We then consider the case when $\Psi_{A\cap B}= \emptyset$. If nodes $\Psi_A$ and $\Psi_B$ are connected in $\Psi_{(A\cup B)^\mathsf{c}}$-cut-sub-graph $G_{\text{cut}(\Psi_{(A\cup B)^\mathsf{c}})}$, then there must be at least one edge $\psi_1\rightarrow \psi_2$ where $\psi_1\in \Psi_A$ and $\psi_2\in \Psi_B$ (or $\psi_1\in \Psi_B$ and $\psi_2\in \Psi_A$), because there is no edge between $\Psi_{A\cup B}$ and $\Psi_{(A\cup B)^\mathsf{c}}$ in $G_{\text{cut}(\Psi_{(A\cup B)^\mathsf{c}})}$. Because $\psi_1\in \Psi_A$, then there must be a directed path $\psi_1\rightsquigarrow x_a$ where $x_a\in X_A^\ast$ by Rule 1. Similarly, because $\psi_2\in \Psi_B$, there must be a directed path $\psi_2\rightsquigarrow x_b$ where $x_b\in X_B^\ast$ by Rule 1. Now we have that there are both $\psi_1\rightsquigarrow x_a$ and $\psi_1\rightsquigarrow x_b$, then by Rule 1 we must include this $\psi_1$ into $\Psi_{A\cap B}$ which leads to a contradiction.
\end{enumerate}

\subsection{Proof of Lemma 2}
By definition $\text{pa}(\Theta_{(A\cup B)^\mathsf{c}})\subseteq \Theta_{A\cup B} \,\cup\, X$, and by \thmcrossref{the first statement of Lemma 1} $\text{ch}(\Theta_{(A\cup B)^\mathsf{c}})=\emptyset$. This implies that $\Theta_{(A\cup B)^\mathsf{c}}$ is independent from $\left(\Theta_{A\cup B},X\right)\setminus \text{pa}(\Theta_{(A\cup B)^\mathsf{c}})$ conditional on $\text{pa}(\Theta_{(A\cup B)^\mathsf{c}})$.
Hence, the standard posterior distribution is
\[
p(\Theta\given X)=p(\Theta_{(A\cup B)^\mathsf{c}}\given \text{pa}(\Theta_{(A\cup B)^\mathsf{c}}))p(\Theta_{A\cup B}\given X).
\]
Since $X = (X_{A\setminus B}, X_{A\cap B}, X_{B\setminus A})$ by the \thmcrossref{first statement of Lemma 1}, we can factorise the second term as
\begin{equation*}
p(\Theta_{A\cup B}\given X) = p(\Theta_{B\setminus A}\given \Theta_{A\setminus B},\Theta_{A\cap B},X_{A\setminus B},X_{A\cap B},X_{B\setminus A})p(\Theta_{A}\given X).
\end{equation*}
We now aim to show:
\[
\Theta_{B\setminus A} \indep (\Theta_{A\setminus B},X_{A\setminus B}) \given  (\Theta_{A\cap B},X_{A\cap B},X_{B\setminus A}).
\]
We prove it by contradiction. Suppose $\Theta_{B\setminus A}$ and $(\Theta_{A\setminus B},X_{A\setminus B})$ are d-connected by $(\Theta_{A\cap B},X_{A\cap B},X_{B\setminus A})$ in DAG $G$. This indicates that there exists an undirected path $U$: $\theta_{B\setminus A}=\psi_1,\cdots,\psi_s=\psi_{A\setminus B}$ between $\Theta_{B\setminus A}$ and $(\Theta_{A\setminus B},X_{A\setminus B})$ such that for every collider $\psi_l$ on this path $U$, either $\psi_l$ or a descendent of $\psi_l$ is in $(\Theta_{A\cap B},X_{A\cap B},X_{B\setminus A})$, and no non-collider on this path $U$ is in $(\Theta_{A\cap B},X_{A\cap B},X_{B\setminus A})$. By the \thmcrossref{second statement of Lemma 1}, we conclude that there is no edge that links $\Theta_{B\setminus A}$ and $(\Theta_{A\setminus B},X_{A\setminus B})$. Hence, all undirected paths that link $\Theta_{B\setminus A}$ and $(\Theta_{A\setminus B},X_{A\setminus B})$ go through nodes in $\Theta_{A\cap B}\cup X_{B\setminus A}\cup X_{A\cap B}\cup \Psi_{(A\cup B)^\mathsf{c}}$.

If path $U$ does not involve any node from $\Psi_{(A\cup B)^\mathsf{c}}$, because of the $d-connection$ by $(\Theta_{A\cap B},X_{A\cap B},X_{B\setminus A})$, there must be a V-structure $\psi_{l-1}\rightarrow\psi_l\leftarrow\psi_{l+1}$ on path $U$ where $\psi_l\in (\Theta_{A\cap B},X_{A\cap B},X_{B\setminus A})$, $\psi_{l-1}\in \Theta_{B\setminus A}$ and $\psi_{l+1}\in (\Theta_{A\setminus B},X_{A\setminus B})$. By the \thmcrossref{second statement of Lemma 1}, there is no edge between $X_{B\setminus A}$ and $(\Theta_{A\setminus B},X_{A\setminus B})$, so $\psi_l\in (\Theta_{A\cap B},X_{A\cap B})$. However, this V-structure has contradicted the \thmcrossref{third statement of Lemma 1}.

If path $U$ involves nodes from $\Psi_{(A\cup B)^\mathsf{c}}$, by the \thmcrossref{first statement of Lemma 1}, there must be a fragment of path $U$: $a=\psi_{s_1},\psi_{s_1+1},\cdots,\psi_{s_2}=b$, $s_1\geq 1$ and $s_2\leq s$, that satisfies $(\psi_{s_1},\psi_{s_2})\in \Psi_{(A\cup B)}$, $(\psi_{s_1+1},\cdots,\psi_{s_2-1})\in \Psi_{(A\cup B)^\mathsf{c}}$, $\psi_{s_1}\rightarrow \psi_{s_1+1}$ and $\psi_{s_2}\rightarrow \psi_{s_2-1}$. Hence, there must be a V-structure $\psi_{l-1}\rightarrow\psi_l\leftarrow\psi_{l+1}$ on this fragment. Because of the $d-connection$ by $(\Theta_{A\cap B},X_{A\cap B},X_{B\setminus A})$, a descendent of $\psi_l\in \Psi_{(A\cup B)^\mathsf{c}}$ is in $(\Theta_{A\cap B},X_{A\cap B},X_{B\setminus A})$. This has contradicted the fact from the \thmcrossref{first statement of Lemma 1} that $\text{ch}(\Psi_{(A\cup B)^\mathsf{c}})=\emptyset$ (i.e., a descendent of $\psi_l\in \Psi_{(A\cup B)^\mathsf{c}}$ must be in $\Psi_{(A\cup B)^\mathsf{c}}$).

In summary, we have proved that the $d-connection$ between $\Theta_{B\setminus A}$ and $(\Theta_{A\setminus B},X_{A\setminus B})$ by $(\Theta_{A\cap B},X_{A\cap B},X_{B\setminus A})$ does not hold. Hence, $\Theta_{B\setminus A}$ and $(\Theta_{A\setminus B},X_{A\setminus B})$ are d-separated by $(\Theta_{A\cap B},X_{A\cap B},X_{B\setminus A})$ and we have
\[
\Theta_{B\setminus A} \indep (\Theta_{A\setminus B},X_{A\setminus B}) \given  (\Theta_{A\cap B},X_{A\cap B},X_{B\setminus A}).
\]
and therefore
\[
p(\Theta_{A\cup B}\given X) = p(\Theta_{B\setminus A}\given \Theta_{A\cap B},X_{A\cap B},X_{B\setminus A})p(\Theta_{A}\given X).
\]
The second result follows by symmetry of A and B.

\subsection{Proof of Theorem 1}
We prove the Theorem for $\Psi_A$: the proof for $\Psi_B$ is the same due to symmetry. Given $X_A^\ast$, Rule 1 indicates that the $X_A^\ast$-associated observables $X_{X_A^\ast} = X_{A\cap B}\cap X_B^\ast$ and the $X_A^\ast$-associated parameters $\Theta_{X_A^\ast} = \Theta_A$ (the parameter of interest to infer the true data generating process of $X_A^\ast$). By \thmcrossref{Definition 2}, we need to build a posterior distribution $p(\Theta_A\given X_A^\ast, X_{A\cap B}\cap X_B^\ast)$ in order to prove $\Phi_A=(\Theta_A,X_A)$ is a minimally self-contained Bayesian module with respect to $X_A^\ast$.

Given the DAG in which $\Psi_A$ and $\Psi_B$ are formed via Rule 1, by \thmcrossref{Lemma 1}, we can write the distribution of each set of variables as follows:
\[
\begin{aligned}
&p(\Theta_{A\setminus B}\given \text{pa}(\Theta_{A\setminus B}));\ \ \text{pa}(\Theta_{A\setminus B}) \subseteq \{\Theta_{A\cap B},X_{A\setminus B},X_{A\cap B}\}; \\
&p(\Theta_{A\cap B}\given \text{pa}(\Theta_{A\cap B}));\ \  \text{pa}(\Theta_{A\cap B}) \subseteq X_{A\cap B} \\
&p(X_{A\setminus B}\given \text{pa}(X_{A\setminus B}));\ \ \text{pa}(X_{A\setminus B}) \subseteq \{\Theta_{A\setminus B},\Theta_{A\cap B},X_{A\cap B}\} \\
&p(X_{A\cap B}\given \text{pa}(X_{A\cap B}));\ \ \text{pa}(X_{A\cap B}) \subseteq \{\Theta_{A\cap B},\Theta_{A\setminus B},\Theta_{B\setminus A},X_{A\setminus B},X_{B\setminus A}\}\\
&p(X_{B\setminus A}\given \text{pa}(X_{B\setminus A}));\ \  \text{pa}(X_{B\setminus A}) \subseteq \{\Theta_{B\setminus A},\Theta_{A\cap B},X_{A\cap B}\} \\
&p(\Theta_{B\setminus A}\given \text{pa}(\Theta_{B\setminus A}));\ \  \text{pa}(\Theta_{B\setminus A}) \subseteq \{\Theta_{A\cap B},X_{B\setminus A},X_{A\cap B}\} \\
&p(\Theta_{(A\cup B)^\mathsf{c}}\given \text{pa}(\Theta_{(A\cup B)^\mathsf{c}}));\ \  \text{pa}(\Theta_{(A\cup B)^\mathsf{c}}) \subseteq \{\Theta_{A\cup B}, X\}.
\end{aligned}
\]
Note that, because $X=X_A^\ast \cup X_B^\ast$, the distribution of $X_{A\cap B}$ can be split into
\begin{align*}
&p(X_{A\cap B}\given \text{pa}(X_{A\cap B}))  = p(X_{A\cap B}\cap X_A^\ast\given \text{pa}(X_{A\cap B}\cap X_A^\ast))p(X_{A\cap B}\cap X_B^\ast\given \text{pa}(X_{A\cap B}\cap X_B^\ast)).
\end{align*}

Because $\text{pa}(X_A^\ast)\cap \Psi_{B\setminus A}=\emptyset$ and $\text{pa}(X_B^\ast)\cap \Psi_{A\setminus B}=\emptyset$, we can write
\[
\begin{aligned}
\text{pa}(X_{A\cap B}\cap X_A^\ast) &\subseteq \{\Theta_{A\cap B},\Theta_{A\setminus B}, X_{A\setminus B},X_{A\cap B}\cap X_B^\ast\} \\
\text{pa}(X_{A\cap B}\cap X_B^\ast) &\subseteq \{\Theta_{A\cap B},\Theta_{B\setminus A}, X_{B\setminus A},X_{A\cap B}\cap X_A^\ast\}.
\end{aligned}
\]

Now given these distributions, one posterior distribution for $\Theta_A$ is simply the standard posterior distribution using all observables. That is:
\[
p(\Theta_A\given X)=\iint p(\Theta_A,\Theta_{B\setminus A},\Theta_{(A\cup B)^\mathsf{c}}\given X) \ d\Theta_{B\setminus A}\ d\Theta_{(A\cup B)^\mathsf{c}}.
\]
This posterior distribution does not justify module A being a self-contained Bayesian module because the posterior involves variables $X_{B\setminus A}$ that are not in module A. However, as we discussed in Section 2.2 of the main text, there could be alternative posterior distributions that do not use all observables. Now we select the following four distributions to build the posterior distribution $p(\Theta_A\given X_A^\ast, X_{A\cap B}\cap X_B^\ast)$. Note that, this selection contains all distributions that do not involve any variable from $X_{B\setminus A}$. The joint distribution is:
\begin{align*}
p(\Theta_{A\setminus B}\given \text{pa}(\Theta_{A\setminus B}))p(\Theta_{A\cap B}\given \text{pa}(\Theta_{A\cap B}))p(X_{A\setminus B}\given \text{pa}(X_{A\setminus B}))p(X_{A\cap B}\cap X_A^\ast\given \text{pa}(X_{A\cap B}\cap X_A^\ast)),
\end{align*}
and it equals to
\begin{equation*}
p(\Theta_{A\setminus B},\Theta_{A\cap B},X_{A\setminus B}, X_{A\cap B}\cap X_A^\ast \given \text{pa}\left(\{\Theta_{A\setminus B},\Theta_{A\cap B},X_{A\setminus B}, X_{A\cap B}\cap X_A^\ast\}\right)).
\end{equation*}
Now we consider $\text{pa}\left(\{\Theta_{A\setminus B},\Theta_{A\cap B},X_{A\setminus B}, X_{A\cap B}\cap X_A^\ast\}\right)$. Since:
\begin{align*}
\text{pa}\left(\{\Theta_{A\setminus B},X_{A\setminus B}\}\right)&\subseteq \{\Theta_{A\cap B},X_{A\cap B}\}; \\
\text{pa}(\Theta_{A\cap B}) &\subseteq X_{A\cap B},
\end{align*}
then
\[
\text{pa}\left(\{\Theta_{A\setminus B},\Theta_{A\cap B}, X_{A\setminus B}\}\right)\subseteq X_{A\cap B} = \{X_{A\cap B}\cap X_A^\ast, X_{A\cap B}\cap X_B^\ast\}.
\]
By Rule 1, a parent node of $X_A^\ast$ belongs to module $A$, so:
\[
\text{pa}(X_{A\cap B}\cap X_A^\ast) \subseteq \{\Theta_{A\setminus B},\Theta_{A\cap B},X_{A\setminus B},X_{A\cap B}\cap X_B^\ast\}.
\]
This implies that
\begin{equation}
\label{eqn:pa1}
\text{pa}\left(\{\Theta_{A\setminus B},\Theta_{A\cap B},X_{A\setminus B}, X_{A\cap B}\cap X_A^\ast\}\right) \subseteq X_{A\cap B}\cap X_B^\ast.
\end{equation}
Now we consider $X_{A\cap B}\cap X_B^\ast$. Given any $a\in X_{A\cap B}\cap X_B^\ast$, by Rule 1, there is at least one directed path $a \rightsquigarrow b$ with a path $a=\psi_1,\psi_2,\cdots,\psi_s=b$, where $b\in X_A^\ast$ and $\psi_i\notin (\Theta_{(A\cup B)^\mathsf{c}},\Theta_{B\setminus A},X_{B\setminus A}\cup X_B^\ast)$ for $i\geq 2$. Hence,
\[
\psi_2 \in \{\Theta_{A\setminus B},\Theta_{A\cap B},X_{A\setminus B}, X_{A\cap B}\cap X_A^\ast\}.
\]
Therefore, we have
\begin{equation}
\label{eqn:pa2}
X_{A\cap B}\cap X_B^\ast \subseteq \text{pa}\left(\{\Theta_{A\setminus B},\Theta_{A\cap B},X_{A\setminus B}, X_{A\cap B}\cap X_A^\ast\}\right).
\end{equation}
Combining \eqref{eqn:pa1} and \eqref{eqn:pa2} implies that
\[
\text{pa}\left(\{\Theta_{A\setminus B},\Theta_{A\cap B},X_{A\setminus B}, X_{A\cap B}\cap X_A^\ast\}\right) = X_{A\cap B}\cap X_B^\ast,
\]
and the joint distribution can be written as:
\[
p\left(\Theta_{A\setminus B},\Theta_{A\cap B},X_{A\setminus B}, X_{A\cap B}\cap X_A^\ast \given X_{A\cap B}\cap X_B^\ast \right).
\]
Note that, $X_A^\ast=X_{A\setminus B}\cup \left(X_{A\cap B}\cap X_A^\ast\right)$. Then we can write a  posterior distribution of $\Theta_A$ as
\begin{equation*}
p(\Theta_A\given X_A^\ast, X_{A\cap B}\cap X_B^\ast) =\frac{p\left(\Theta_{A\setminus B},\Theta_{A\cap B},X_{A\setminus B},X_{A\cap B}\cap X_A^\ast\given X_{A\cap B}\cap X_B^\ast\right)}{\int p\left(\Theta_{A\setminus B},\Theta_{A\cap B},X_{A\setminus B},X_{A\cap B}\cap X_A^\ast\given X_{A\cap B}\cap X_B^\ast\right) d\Theta_A},
\end{equation*}
where $X_A^\ast=X_{A\setminus B}\cup \left(X_{A\cap B}\cap X_A^\ast\right)$ and $X_{supp}=\emptyset$, by \thmcrossref{Definition 2}, we conclude that $\Psi_A=(\Theta_A,X_A)$ forms a minimally self-contained Bayesian module with respect to observable random variables $X_A^\ast$.

\subsection{Proof of Lemma 3}
We give a proof by contraction. Suppose $\Psi_{A\setminus B}$ and $\Psi_{B\setminus A}$ are d-connected by $\Psi_{A\cap B}$. This indicates that there exists an undirected path $U$: $\psi_{A\setminus B}=\psi_1,\cdots,\psi_s=\psi_{B\setminus A}$ between $\Psi_{A\setminus B}$ and $\Psi_{B\setminus A}$ such that for every collider $\psi_l$ on this path $U$, either $\psi_l$ or a descendent of $\psi_l$ is in $\Psi_{A\cap B}$, and no non-collider on this path $U$ is in $\Psi_{A\cap B}$. By the \thmcrossref{second statement of Lemma 1}, we conclude that there is no edge that links $\Psi_{A\setminus B}$ and $\Psi_{B\setminus A}$. Hence, all undirected paths that link $\Psi_{A\setminus B}$ and $\Psi_{B\setminus A}$ go through nodes in $\Psi_{A\cap B}\cup \Psi_{(A\cup B)^\mathsf{c}}$.

If path $U$ does not involve any node from $\Psi_{(A\cup B)^\mathsf{c}}$, because of the $d-connection$ by $\Psi_{A\cap B}$, there must be a V-structure $\psi_{l-1}\rightarrow\psi_l\leftarrow\psi_{l+1}$ on path $U$ where $\psi_l\in \Psi_{A\cap B}$, $\psi_{l-1}\in \Psi_{A\setminus B}$ and $\psi_{l+1}\in \Psi_{B\setminus A}$. This has contradicted the \thmcrossref{third statement of Lemma 1}.

If path $U$ involves nodes from $\Psi_{(A\cup B)^\mathsf{c}}$, by the \thmcrossref{first statement of Lemma 1}, there must be a fragment of path $U$: $a=\psi_{s_1},\psi_{s_1+1},\cdots,\psi_{s_2}=b$, $s_1\geq 1$ and $s_2\leq s$, that satisfies $(\psi_{s_1},\psi_{s_2})\in \Psi_{(A\cup B)}$, $(\psi_{s_1+1},\cdots,\psi_{s_2-1})\in \Psi_{(A\cup B)^\mathsf{c}}$, $\psi_{s_1}\rightarrow \psi_{s_1+1}$ and $\psi_{s_2}\rightarrow \psi_{s_2-1}$. Hence, there must be a V-structure $\psi_{l-1}\rightarrow\psi_l\leftarrow\psi_{l+1}$ on this fragment. Because of the $d-connection$ by $\Psi_{A\cap B}$, a descendent of $\psi_l\in \Psi_{(A\cup B)^\mathsf{c}}$ is in $\Psi_{A\cap B}$. This has contradicted the fact that $\text{ch}(\Psi_{(A\cup B)^\mathsf{c}})=\emptyset$ (i.e., a descendent of $\psi_l\in \Psi_{(A\cup B)^\mathsf{c}}$ must be in $\Psi_{(A\cup B)^\mathsf{c}}$).

In summary, we have proved that the $d-connection$ between $\Psi_{A\setminus B}$ and $\Psi_{B\setminus A}$ by $\Psi_{A\cap B}$ does not hold. Hence, $\Psi_{A\setminus B}$ and $\Psi_{B\setminus A}$ are d-separated by $\Psi_{A\cap B}$ and we have
\[
\Psi_{A\setminus B}\indep \Psi_{B\setminus A} \given  \Psi_{A\cap B}.
\]

When the modules are unordered neither $A\rightharpoonup B$ nor $B\rightharpoonup A$ hold, so $\text{ch}(\Psi_{A \cap B}) \cap \Psi_{B\setminus A} = \emptyset$ and $\text{ch}(\Psi_{A \cap B}) \cap \Psi_{A\setminus B} = \emptyset$, which implies that $\text{ch}(\Psi_{A \cap B}) \subseteq \Psi_{(A \cup B)^\mathsf{c}}$
We proceed by contradiction. Suppose $\Psi_{A \cap B} \neq \emptyset$, so there must exist a node $\psi \in \Psi_{A \cap B}$. 
Then there must exist a directed path $\psi=\psi_{1},\cdots,\psi_{s} \in \Psi_{A \setminus B}$. Since $\text{ch}(\Psi_{A \cap B}) \subseteq \Psi_{(A \cup B)^\mathsf{c}}$, $\psi_{2} \in \Psi_{A \cap B} \cup \Psi_{(A \cup B)^\mathsf{c}}$, so $\Psi_{2} \nsubseteq \Psi_{A\setminus B}$ and thus $s >2$. Since $\psi_{s} \notin \Psi_{A \cap B}$ there must exist a $j \in \{2, \dots, s\}$ such that $\psi_{j-1} \in \Psi_{A \cap B}$ and $\psi_{j} \notin \Psi_{A \cap B}$. Now $\text{ch}(\psi_{j-1}) \in \Psi_{(A \cup B)^\mathsf{c}}$, so $\psi_{j}, \cdots, \psi_{s} \in \Psi_{(A \cup B)^\mathsf{c}}$ by \thmcrossref{statement 1 of Lemma 1}. This contradicts $\psi_{s} \in \Psi_{A \setminus B}$, so $\Psi_{A \cap B} = \emptyset$.

\subsection{Proof of Theorem 2}
Because both the joint distribution $p(X,\Theta)$ and $p_f(\Theta)$ involve the term $p(\Theta_{(A\cup B)^\mathsf{c}}\given \text{pa}(\Theta_{(A\cup B)^\mathsf{c}}))$, it is cancelled in the denominator and numerator of the logarithm term of KL divergence. Therefore, by denoting the marginal joint distribution
\[
p(\Theta_A,\Theta_{B\setminus A},X) = \frac{p(X,\Theta)}{p(\Theta_{(A\cup B)^\mathsf{c}}\given \text{pa}(\Theta_{(A\cup B)^\mathsf{c}}))},
\]
the KL divergence is
\begin{equation*}
\mathbb{D}_{KL}\left(p_f(\cdot),p(X,\cdot)\right) = \int f(\Theta_{B\setminus A}) p(\Theta_A\given X_A) \log\frac{f(\Theta_{B\setminus A}) p(\Theta_A\given X_A)}{p(\Theta_A,\Theta_{B\setminus A},X)} d\Theta_A d\Theta_{B\setminus A}.
 \end{equation*}
By the \thmcrossref{proof of Theorem 1}, we have
\[
\begin{aligned}
p(\Theta_A\given X_A) = \frac{p(\Theta_{A\setminus B},\Theta_{A\cap B},X_{A\setminus B},X_{A\cap B}\cap X_A^\ast\given X_{A\cap B}\cap X_B^\ast)}{C},
\end{aligned}
\]
where the numerator is a product of $p(\Theta_{A\setminus B}\given \text{pa}(\Theta_{A\setminus B}))$, $p(\Theta_{A\cap B}\given \text{pa}(\Theta_{A\cap B}))$, $p(X_{A\setminus B}\given \text{pa}(X_{A\setminus B}))$ and $p(X_{A\cap B}\cap X_A^\ast\given \text{pa}(X_{A\cap B}\cap X_A^\ast))$ which are also involved in the marginal joint distribution $p(\Theta_A,\Theta_{B\setminus A},X)$ and $C$ is a constant that does not depend on any component of the parameter $\Theta$. Therefore, we can cancel the numerator of $p(\Theta_A\given X_A)$ with the corresponding terms in $p(\Theta_A,\Theta_{B\setminus A},X)$ in the logarithm term of the KL divergence and then integral $\Theta_A$. The KL divergence is reduced to
\begin{equation*}
\mathbb{D}_{KL}\left(p_f(\cdot),p(X,\cdot)\right)= \int f(\Theta_{B\setminus A})\log \frac{ f(\Theta_{B\setminus A}) \frac{1}{C}}{p(\Theta_{B\setminus A},X_{B\setminus A},X_{A\cap B}\cap X_B^\ast\given \text{pa}(\{\Theta_{B\setminus A},X_{B\setminus A},X_{A\cap B}\cap X_B^\ast\}))}d\Theta_{B\setminus A}
\end{equation*}
To minimize this KL divergence, we require:
\begin{equation*}
f(\Theta_{B\setminus A}) \propto p(\Theta_{B\setminus A},X_{B\setminus A},X_{A\cap B}\cap X_B^\ast\given \text{pa}(\{\Theta_{B\setminus A},X_{B\setminus A},X_{A\cap B}\cap X_B^\ast\})).
\end{equation*}
We now consider
\begin{equation*}
p(\Theta_{B\setminus A},X_{B\setminus A},X_{A\cap B}\cap X_B^\ast\given \text{pa}(\{\Theta_{B\setminus A},X_{B\setminus A},X_{A\cap B}\cap X_B^\ast\})).
\end{equation*}
By the \thmcrossref{proof of Theorem 1}, we have
\begin{align*}
\text{pa}(\{\Theta_{B\setminus A},X_{B\setminus A}\})&\subseteq \{\Theta_{A\cap B},X_{A\cap B}\}; \\
\text{pa}(X_{A\cap B}\cap X_B^\ast)&\subseteq \{\Theta_{A\cap B},\Theta_{B\setminus A}, X_{B\setminus A},X_{A\cap B}\cap X_A^\ast\}.
\end{align*}
Hence,
\[
\text{pa}(\{\Theta_{B\setminus A},X_{B\setminus A},X_{A\cap B}\cap X_B^\ast\})\subseteq \{\Theta_{A\cap B}, X_{A\cap B}\cap X_A^\ast\}.
\]
Although the right side term $\{\Theta_{A\cap B}, X_{A\cap B}\cap X_A^\ast\}$ may be larger than the parent nodes of $\{\Theta_{B\setminus A},X_{B\setminus A},X_{A\cap B}\cap X_B^\ast\}$, we can write
\[
f(\Theta_{B\setminus A}) \propto p(\Theta_{B\setminus A},X_{B\setminus A},X_{A\cap B}\cap X_B^\ast\given \Theta_{A\cap B}, X_{A\cap B}\cap X_A^\ast).
\]
Finally, we have
\[
f(\Theta_{B\setminus A}) = p(\Theta_{B\setminus A}\given \Theta_{A\cap B},X_{B\setminus A},X_{A\cap B}).
\]

\subsection{Proof of Lemma 4}
We first consider the case when $\Psi_{\underline{BC}}\neq\emptyset$. By Rule 1, there must be at least one edge $a\rightarrow b$ where $a\in\Psi_{\underline{BC}}$ and $b$ is either from $\Psi_{\underline{B}}\cup \Psi_{\underline{C}}$ or $\underline{\underline{A}}$. We only need to consider the case when $b\in \underline{\underline{A}}$. If $b\in \Theta\cap \underline{\underline{A}}$, then $a\in \Psi_A$ and this has contradicted the fact that $a\in\Psi_{\underline{BC}}$. If $b\in X \cap \underline{\underline{A}}$, then $b$ must not be within $X_A^\ast$ so that $a$ is not in $\Psi_A$. Hence $b\in X_B^\ast\cup X_C^\ast$.

We now consider the case when $\Psi_{\underline{BC}}=\emptyset$. We first prove that if there is no edge that directly links $\underline{B}$ and $\underline{C}$ by contradiction.
\begin{itemize}
\item Without loss of generality, if there is an edge $\psi_b\rightarrow x_c$ where $\psi_b\in\Psi_{\underline{B}}$ and $x_c\in X_{\underline{C}}$, since $X_{\underline{C}}\subseteq X_C^\ast$, then by Rule 1, we have that $\psi_b\in \Psi_C$. Given that $\psi_b\in\Psi_B$, we have $\psi_b\in\Psi_{B\cap C}$. This has contradicted the fact that $\psi_b\in\Psi_{\underline{B}}$.

\item If this edge $\psi_b\rightarrow \theta_c$ satisfies $\psi_b\in\Psi_{\underline{B}}$ and $\theta_c\in \Theta_{\underline{C}}$, by Rule 1, there must be a directed path with an observable random variable $x_c\in X_C^\ast$ as the leaf and $\theta_c$ as the root. By Rule 1 again, we conclude that we must incorporate $\psi_b$ into module $C$. Therefore we have $\psi_b\in\Psi_{B\cap C}$ which contradicts the fact that $\psi_b\in\Psi_{\underline{B}}$.
\end{itemize}

Similarly to the proof of \thmcrossref{Lemma 3}, we now give a proof by contradiction. Suppose $\Psi_{\underline{B}}$ and $\Psi_{\underline{C}}$ are d-connected by $\Psi_A$ in DAG $G$. This indicates that there exists an undirected path $U$: $\psi_{\underline{B}}=\psi_1,\cdots,\psi_s=\psi_{\underline{C}}$ between $\Psi_{\underline{B}}$ and $\Psi_{\underline{C}}$ such that for every collider $\psi_l$ on this path $U$, either $\psi_l$ or a descendent of $\psi_l$ is in $\Psi_A$, and no non-collider on this path $U$ is in $\Psi_A$. Because there is no edge that links $\Psi_{\underline{B}}$ and $\Psi_{\underline{C}}$, all undirected paths that link $\Psi_{\underline{B}}$ and $\Psi_{\underline{C}}$ go through nodes in $\Psi_A\cup \Psi_S$.

If path $U$ does not involve any node from $\Psi_S$, because of the $d-connection$ by $\Psi_A$, there must be a V-structure $\psi_{l-1}\rightarrow\psi_l\leftarrow\psi_{l+1}$ on path $U$ where $\psi_l\in \Psi_A$, $\psi_{l-1}\in \Psi_{\underline{B}}$ and $\psi_{l+1}\in \Psi_{\underline{C}}$. Because $\psi_l\in\Psi_A$, we must include $\psi_{l-1}$ and $\psi_{l+1}$ into $\Psi_A$ according to Rule $1$, this has contradicted the fact that $\Psi_{\underline{B}}\cup \Psi_{\underline{C}}$ does not intersect with $\Psi_A$.

If path $U$ involves nodes from $\Psi_S$, by the \thmcrossref{first statement of Lemma 1}, there must be a fragment of path $U$: $a=\psi_{s_1},\psi_{s_1+1},\cdots,\psi_{s_2}=b$, $s_1\geq 1$ and $s_2\leq s$, that satisfies $(\psi_{s_1},\psi_{s_2})\in \Psi_{(A\cup B\cup C)}$, $(\psi_{s_1+1},\cdots,\psi_{s_2-1})\in \Psi_S$, $\psi_{s_1}\rightarrow \psi_{s_1+1}$ and $\psi_{s_2}\rightarrow \psi_{s_2-1}$. Hence, there must be a V-structure $\psi_{l-1}\rightarrow\psi_l\leftarrow\psi_{l+1}$ on this fragment. Because of the $d-connection$ by $\Psi_A$, a descendent of $\psi_l\in \Psi_S$ is in $\Psi_A$. This has contradicted the fact that $\text{ch}(\Psi_S)=\emptyset$ (i.e., a descendent of $\psi_l\in \Psi_S$ must be in $\Psi_S$).

In summary, we have proved that the $d-connection$ between $\Psi_{\underline{B}}$ and $\Psi_{\underline{C}}$ by $\Psi_A$ does not hold. Hence, $\Psi_{\underline{B}}$ and $\Psi_{\underline{C}}$ are d-separated by $\Psi_A$ and we have
\[
\Psi_{\underline{B}} \indep \Psi_{\underline{C}} \given  \Psi_A.
\]

\section{Illustration of cut inference on the generic two-module case}
\label{appB}
\begin{figure}[!t]
\setlength{\abovecaptionskip}{0cm}
\setlength{\belowcaptionskip}{0cm}
\centering
\includegraphics[width=0.5\textwidth]{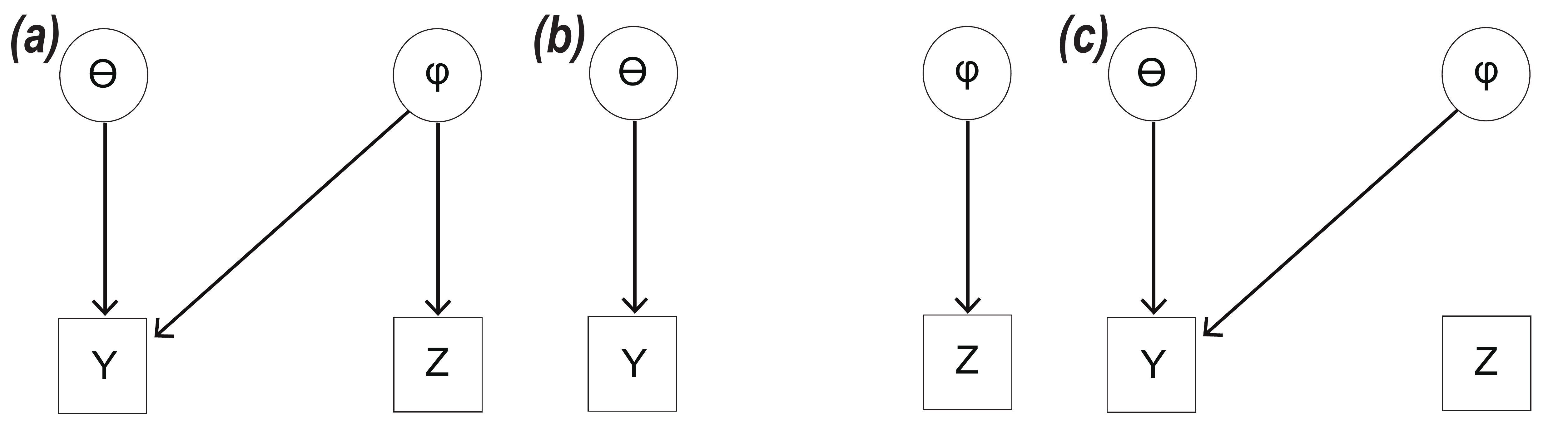}
\caption[DAG representation of a generic two-module model.]{\textbf{DAG representation of a generic two-module model.} (a). The original DAG $G$ for the joint distribution. (b). The cut-sub-graph $G_{\text{cut}(\Psi_{(B\setminus A)})}$. (c). The cut-sub-graph $G_{\text{cut}(\Psi_{(A\setminus B)})}$. The modules are $\Psi_A=(\varphi,Z)$ and $\Psi_B=(\theta,\varphi,Y)$.}
\label{F8}
\end{figure}

We consider the most commonly-used and simplest two-module case, which has been extensively previously considered \parencite[e.g.,][]{plummer2015cuts,https://doi.org/10.1111/rssb.12336,carmona2020semi,liu2020stochastic}. Suppose that we have observable random variables $Y$ and $Z$ and parameters $\varphi$ and $\theta$, as shown in Figure \ref{F8} (a). The joint distribution, given the likelihood and prior distributions, is
\[
p(Y,Z,\theta,\varphi)=p(Y\given \theta,\varphi)p(Z\given \varphi)p(\theta)p(\varphi),
\]
and the standard posterior distribution is:
\begin{equation}
    p(\theta,\varphi\given Y,Z) = p(\theta\given \varphi,Y)p(\varphi\given Y,Z).
\label{E12}
\end{equation}

We first partition the observable random variables into two disjoint groups. Suppose we choose the following split: $X_A^\ast=Z\in \Psi_A$ and $X_B^\ast=Y\in \Psi_B$. By Rule 1, we then enlarge $\Psi_A$ and $\Psi_B$ by incorporating the ancestors of $Z$ and $Y$. This leads to two self-contained Bayesian modules: $\Psi_A=(\varphi,Z)$ and $\Psi_B=(\theta,\varphi,Y)$. It can be easily checked that these are self-contained Bayesian modules because the following two posterior distributions are well-defined:

\begin{subequations}
\begin{align}
p_A(\varphi\given Z) &\propto p(Z\given \varphi)p(\varphi); \label{E13:subeq1}\\
p_B(\theta,\varphi\given Y) &\propto p(Y\given \theta,\varphi)p(\theta)p(\varphi). \label{E13:subeq2}
\end{align}
\end{subequations}

Next we identify the order between modules $A$ and $B$. There are two directed edges from $\Psi_{A\cap B}=\varphi$ to $\Psi_{B\setminus A}=(\theta,Y)$ or $\Psi_{A\setminus B}=Z$: $\varphi\rightarrow Y$ and $\varphi \rightarrow Z$. Thus, by Rule 2, either module $A$ or $B$ can be the parent module. The assignment of the parent module typically depends on how much we trust the information provided each module.

Suppose we know that there is a misspecification of module $B$ and consequently we want to prevent inference of $\varphi$ being affected by information from module $B$. In this case, we should regard module $A$ as the parent module. To cut the feedback, we first obtain the cut-sub-graph $G_{\text{cut}(\Psi_{(B\setminus A)})}$ as shown in Figure \ref{F8} (b). Then by Rule 3, we can infer $\varphi$ by only using information from module $A$ by using its marginal posterior \eqref{E13:subeq1}. To infer the child module $B$, we use conditional self-contained Bayesian inference, conditioning on $\varphi$ for the distribution of $\theta$. The cut distribution is
\[
p_{A\rightharpoonup B}(\theta,\varphi) = p(\theta\given \varphi,Y)p_A(\varphi\given Z) \propto p(Y\given \theta,\varphi)p_A(\varphi\given Z)p(\theta).
\]
It is clear that the inference of module $A$ (i.e., $\varphi$) is not affected by $\Psi_{B\setminus A}$.

Suppose now that we wish to regard module $B$ as the parent module so that $B\rightharpoonup A$ . In this case, the cut-sub-graph $G_{\text{cut}(\Psi_{(A\setminus B)})}$, shown in Figure \ref{F8} (c), is needed to apply Rule 3. Notice that cutting the feedback here results in the observable random variable $Z$ not being used: the cut distribution is simply the posterior distribution of module $B$ as \eqref{E13:subeq2}:
\[
p_{B\rightharpoonup A}(\theta,\varphi) = p_B(\theta,\varphi\given Y).
\]

Comparing $p_{A\rightharpoonup B}(\theta,\varphi)$ and $p_{B\rightharpoonup A}(\theta,\varphi)$ with the standard posterior distribution $p(\theta,\varphi\given Y,Z)$, we conclude that no matter which module is regarded as the parent module, the parent module is always inferred without being affected by information from the child module. This is in contrast to the standard posterior distribution \eqref{E12}, under which $\varphi$ depends on both $Y$ and $Z$ and $\theta$ depends on $Y$ and $\varphi$, meaning that inference is a mixture of information from both module $A$ and module $B$.

\section{Numerical simulation of section 3.2}
\label{appC}
Suppose the true data generating process for observations is equation (6) in the main text, where $(a_1,\cdots,a_T)$ is set as 0 for simplification and the elements of covariates $(p_{t1},\cdots,p_{tn})$ and $(q_{t1},\cdots,q_{tn})$, $t= 1,2,\cdots,100$ are independently generated from normal distribution with mean 0. At times $t= 1,2,\cdots,100$, suppose the true value for the parameter is $\theta_t=10\sin(t)$, and the ``linking'' function $f^\ast(\cdot)$ in the true data generating process is $f^\ast(\theta) = \theta$.

Here, we consider the following three scenarios for the misspecification of the function $f(\cdot)$ that we use in the model:
\[
\begin{aligned}
& f(\theta) = f^\ast(\theta)-2 = \theta - 2;\ \ \ \textbf{Upper biased}; \\
& f(\theta) = f^\ast(\theta)+2 = \theta + 2;\ \ \ \textbf{Lower biased};\\
& f(\theta) = f^\ast(\theta)= \theta ;\ \ \ \textbf{Not biased};
\end{aligned}
\]
We simulate $n=100$ observations independently according to the true data generating process at each $t$. Hence, we have observations $X_t=(X_{t1},\cdots,X_{t100})$, $t=1,\cdots,100$.

\begin{figure}[!htb]
\setlength{\abovecaptionskip}{0cm}
\setlength{\belowcaptionskip}{0cm}
\centering
\includegraphics[width=.5\textwidth]{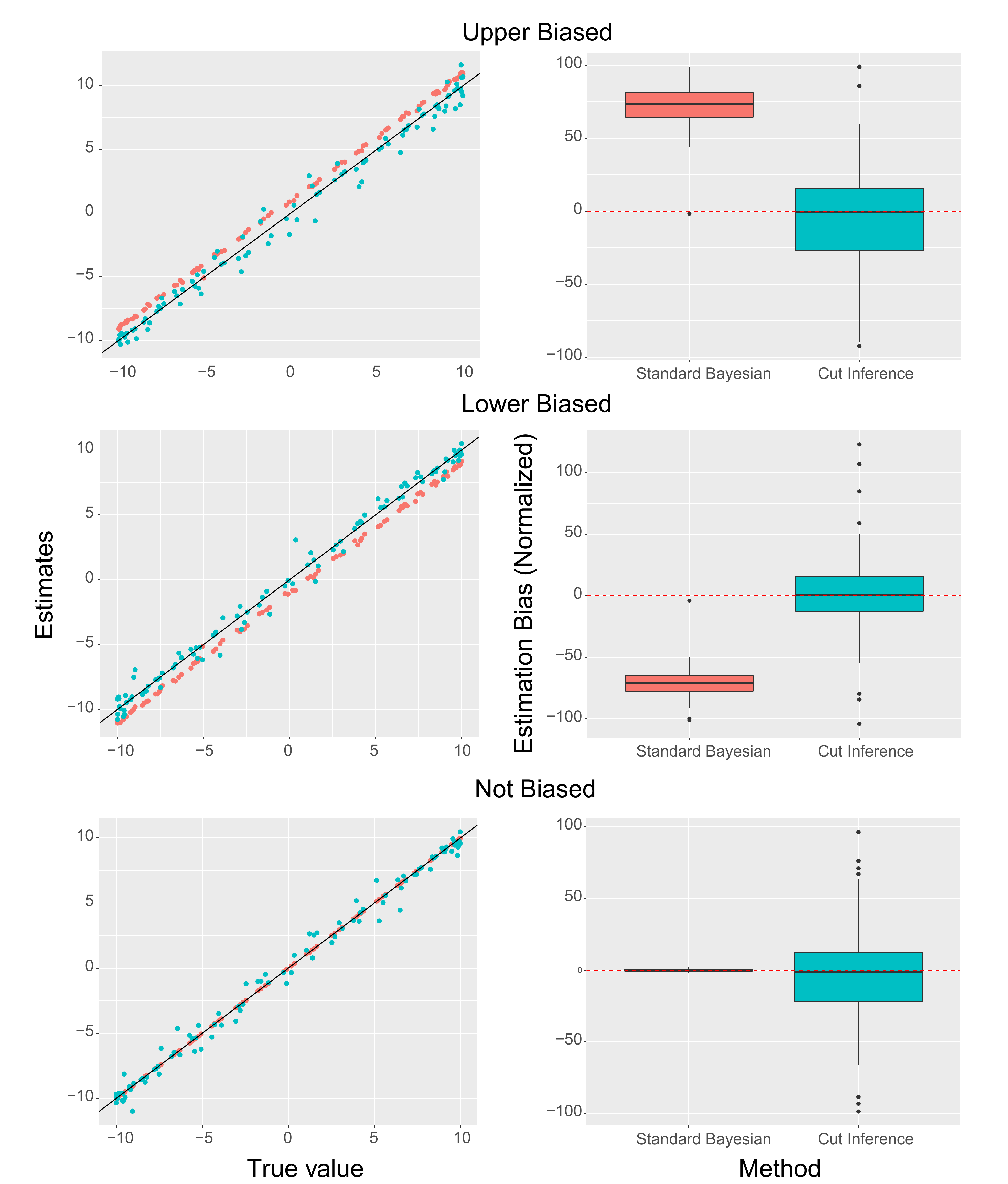}
\caption[Numerical simulation results of cut inference and standard Bayesian inference under a multiple-module case.]{\textbf{Numerical simulation results of cut inference and standard Bayesian inference under a multiple-module case.} Left: scatter plots of the estimates verses true value of parameters. Right: boxplots of the estimation bias after normalization. The estimation bias is normalized according to the standard deviation of the estimates.}
\label{F11}
\end{figure}

We now consider and compare the cut inference with the standard Bayesian inference under three scenarios. We use a Metropolis-Hasting algorithm for standard Bayesian inference and WinBUGS algorithm \parencite{https://doi.org/10.1002/sim.3680} for the cut inference. The WinBUGS algorithm works fine here because the model has a simple linear form. We draw samples from both samplers until the chains converge and use the posterior mean from the samples as our estimate of $\theta_t$, $t=1,\cdots,100$. The results are shown in Figure \ref{F11}.

We first look at the scenarios when there is a deterministic bias. For the upper bias scenario, estimates of standard Bayesian inference all overestimate the parameter. Similarly, estimates of standard Bayesian inference all underestimate the parameter when there is a lower bias instead. In contrast, the estimates using cut inference evenly distribute around the truth. This indicates that cut inference reduces the estimation bias in this simulation, as the theory in Section 3.2 indicated.

We now consider the scenario when there is no misspecification. Both standard Bayesian inference and cut inference have correctly estimated the parameter on an average basis with respect to sampling time. Unlike the scenario of misspecification, standard Bayesian inference outperforms cut inference in terms of the variance of the estimation bias. This is expected as the standard Bayesian inference utilizes all available information when the whole model is correctly specified. In contrast, cut inference has prevented the feedback from descendant modules so information is only partially used to infer the parameter.

\section{Definition of d-separation used in this paper}
\label{appD}
In a DAG, a path between $\psi_1$ and $\psi_m$ $(\psi_1, \psi_2,\cdots, \psi_m)$ is blocked by a set $T$ with neither $\psi_1$ nor $\psi_m$ in $T$ whenever there is a node $\psi_j$ such that one of the following two possibilities hold:
\begin{itemize}
    \item $\psi_j\in T$ and we don't have $\psi_{j-1}\rightarrow \psi_j \leftarrow \psi_{j+1}$.
    \item $\psi_{j-1}\rightarrow \psi_j \leftarrow \psi_{j+1}$ and neither $j_l$ nor any of its descendants are in $T$.
\end{itemize}

If $G$ is a DAG, given a triple of subsets of nodes $A$, $B$, $T$, we say $T$ d-separates $A$ from
$B$ if $T$ blocks every path from $A$ to $B$.


\printbibliography

\end{document}